\documentclass[aps,prc,twocolumn,showpacs,amsmath,amssymb,nofootinbib]{revtex4-1}
\usepackage{graphicx}
\usepackage{color}
\usepackage{times}
\usepackage{inputenc}
\usepackage{bm}
\usepackage{ulem}
\usepackage{multirow}
\usepackage{float}
\usepackage{url}
\usepackage{natbib}
\usepackage{mathrsfs}
\usepackage{physics}
\usepackage{comment}
\usepackage[table,xcdraw]{xcolor}
\usepackage{booktabs}
\usepackage{comment}
\usepackage[caption=false]{subfig}
\usepackage[colorlinks=true,citecolor=blue,urlcolor=blue,linkcolor=blue]{hyperref}
\newcommand{\ols}[1]{\mskip.4\thinmuskip\overline{\mskip-.4\thinmuskip {#1} \mskip-.4\thinmuskip}\mskip.4\thinmuskip} 
\begin{document}
\title{Magnetised neutron star crust within effective relativistic mean-field model}
\author{Vishal Parmar$^{1}$}
\email{physics.vishal01@gmail.com}
\author{H. C. Das$^{2,3}$}
\email{harishdas.physics@gmail.com}
\author{M. K. Sharma$^{1}$}
\author{S. K. Patra$^{2,3}$} 
\affiliation{\it $^{1}$ School of Physics and Materials Science, Thapar Institute of Engineering and Technology, Patiala 147004, India}
\affiliation{\it $^{2}$Institute of Physics, Sachivalaya Marg, Bhubaneswar 751005, India}
\affiliation{\it $^{3}$Homi Bhabha National Institute, Training School Complex, Anushakti Nagar, Mumbai 400094, India}
\date{\today}
\begin{abstract}
Even though the crystallize nature of the neutron star crust plays a pivotal role in describing various fascinating astrophysical observations, its microscopic structure is not fully understood in the presence of a colossal magnetic field. In the present work, we study the crustal properties of a  neutron star within an effective relativistic mean field framework in the presence of magnetic field strength $\sim 10^{17}$G. We calculate the equilibrium composition of the outer crust by minimizing the Gibbs free energy using the most recent atomic mass evaluations. The magnetic field significantly affects the equation of state (EoS) and the properties of the outer crust, such as neutron drip density, pressure, and melting temperature. For the inner crust, we use the compressible liquid drop model for the first time to study the crustal properties in a magnetic environment. The inner crust properties, such as mass and charge number distribution, isospin asymmetry, cluster density, etc., show typical quantum oscillations (De Haas–van Alphen effect) sensitive to the magnetic field's strength. The density-dependent symmetry energy influences the magnetic inner crust like the field-free case. We study the probable modifications  in the pasta structures  and it is observed that their mass and thickness changes by $\sim 10-15 \%$  depending upon the magnetic field strength. The fundamental torsional oscillation mode frequency is investigated for the magnetized crust in the context of quasiperiodic oscillations (QPO) in soft gamma repeaters. The magnetic field strengths considered in this work influences only the EoS of outer and shallow regions of the inner crust, which results in no significant change in global neutron star properties. However, the outer crust mass and its moment of inertia increase considerably with increase in  magnetic field strength. 
\end{abstract}
\maketitle
\section{Introduction}
\label{intro}
In a recent breakthrough, astronomers detected an extremely bright radio burst from the Galactic magnetar SGR 1935+2154 \cite{chime2020bright, bochenek2020fast}, which confirmed that the gamma-ray bursts (GRBs) originate from the magnetars at cosmological distances \cite{kirsten2021detection}. Magnetars are the family of neutron stars with an extremely intense magnetic field ($\ge 10^{15}$G) known for their observed quiescent in a wide range of the electromagnetic spectrum that includes $X-$  ray and $\gamma-$ ray in the form of powerful bursting emissions \cite{Kaspi_2017, Konstantinos_2016}. The origin of the colossal magnetic field in the magnetar is still controvertible; however, a common hypothesis is that strong dynamo effects caused due to the initial spin period are responsible for such an extreme environment \cite{Thompson_1995, Vink_2006}. The activities of the magnetars are principally associated with the crustal motion, which twists its magnetosphere \cite{Beloborodov_2014, Thompson_2002}.
Another class of neutron stars with a strong magnetic field ($10^{12}-10^{14}$ G) are the pulsars \cite{lorimer2008binary} (majority of the observed neutron stars are pulsars), which convert the rotational energy of the star into periodic multi-wavelength radiations \cite{Kaspi_2010, Baiko_2017}. Magnetars, along with pulsars, provide extraordinary opportunities  to develop and test theories or models to describe and explain the wide range of associated observational phenomena \cite{Turolla_2015, mereghetti2015magnetars}, such as gamma-ray bursts (GMR), fast radio bursts (FRB), $X-$ ray outbursts, etc.

In general, the global properties such as mass, radius, the moment of inertia, etc., of a magnetar or pulsar are dictated by their core, where density reaches $\sim 10$ times the nuclear saturation density, and the matter is considered to be homogeneous \cite{Haenel_2007}. The core is covered by $\sim 1$ km thick heterogeneous crust characterized by fully ionized nuclei submerged in strongly degenerate electron gas known as the outer crust and the nuclear clusters (spherical in the shallower region and distorted (nuclear pasta) in dense regions) surrounded by electron and degenerate dripped neutron gas known as inner crust \cite{chamel2008physics}. This layer of the neutron star is of primary interest to nuclear and astrophysicists as it acts as a unique exotic non-terrestrial laboratory to test theories of strong interaction and validate them using various observations. Recently it was shown that the crust plays a crucial role in stabilizing the magnetic field by solidification, which results in the development of elastic forces that consequently avoid the fast decay of the magnetic field \cite{Lander_2021}.  Therefore, an accurate description of neutron star crust in the presence of a magnetic field is essential to extract the core's properties and understand microscopic aspects of  the crust, such as cooling \cite{Ootes_2018}, entrainment,  quasiperiodic oscillations (QPOs) \cite{Stein_2016}, torsional vibrations, shattering \cite{Tsang_2009}, transport \cite{rezzolla2018physics} etc. The astrophysical phenomena related to the interaction and evolution of the magnetic field in the neutron star crust \cite{Baiko_2011, Konstantinos_2016, Sengo_2020} also make the study of neutron star crust in a magnetic environment highly desirable.

There have been numerous attempts in the last few decades to study the neutron star's outer and inner crust since they were first estimated for a cold nonaccreting unmagnetized neutron star by Baym-Pethick-Sutherland (BPS) \cite{BPS_1971} and Baym-Bethe-Pethick (BBP) \cite{Baym_1971} respectively. While the outer crust calculations are relatively straightforward, where the nuclear masses act as input parameters, the inner crust structure is determined by  considering the dripped neutron gas, an environment nonviable for a terrestrial laboratory. The persistent uncertainties in the nuclear matter observables make the inner crust calculations inevitably model-dependent. Various forms of nuclear interaction and inner crust models such as microscopic Hartree-Fock \cite{NEGELE1973298}, Thomas-Fermi (TF) \cite{BKS_2015}, extended TF (ETF) \cite{Baldo_2006, Onsi_2008}, compressible liquid drop model (CLDM) \cite{Parmar_2022_1, Parmar_2022_2, Newton_2013, Carreau_2019}, nuclear density functional theory \cite{schuetrumpf2016clustering}, molecular dynamics simulations \cite{Engstrom_2016, watanabe2007dynamical} etc. have been used in past to investigate the inner crust.  The principal aim of these studies was to perform structural analyses such as elasticity properties to understand the neutron star crust oscillations, equilibrium composition, transition properties, etc., and examine their model dependencies \cite{Sotani_2016, Newton_2012}. The inner crust structure is  dependent on the density dependence of symmetry energy in the subsaturation density regime of the nuclear matter \cite{Pearson_2020, Parmar_2022_1, Parmar_2022_2}. 

Most neutron star crust calculations in literature have been performed for an unmagnetized neutron star, and not much emphasis is given to the magnetized crust. Among a handful of studies that consider the magnetic field, the majority consider only the outer crust composition \cite{Basilico_2015, Chamel_2012, Chamel_2015, Arteaga_2011, Stein_2016}. The inner crust calculations are  either limited to the effect of magnetic field on the electrons  \cite{Mutafchieva_2019} or study  the crust-core transition properties employing Vlasov formalism for dynamical instability \cite{fang_2017, Fang_2017_1}. Only a few inner crust calculations have been performed using the self-consistent Thomas-Fermi approximation employing relativistic mean field theory \cite{Bao_2021, Nandi_2011, Lima_2013}, some of which \cite{Lima_2013} consider a fixed proton fraction instead of $\beta$-equilibrium in the inner crust. Therefore, the lack of comprehensive investigations of magnetized neutron crust in a unified manner using the realistic equation of state (EoS), which satisfies relevant nuclear matter and neutron star constraints, underscores the need for such analysis.  

In this work, we aim to investigate the possible changes in the neutron star crust's structure due to the presence of  magnetic field  and its EoS and study associated phenomena. Recently, we studied the crustal properties of the neutron star \cite{Parmar_2022_1, Parmar_2022_2}, using the CLDM method employing effective relativistic mean field model (E-RMF). We extend these calculations for the case of magnetic fields by incorporating the magnetic field effects in the EoS. This will help us to understand the neutron star crust structure in a magnetic environment comprehensively and analyze the possible deviations as compared to the unmagnetized neutron star. For the first time, we use the CLDM method to estimate the crust structure of a magnetized neutron star. The CLDM method to estimate inner crust structure is known to be efficient and produces results at par with the TF calculations \cite{Newton_2012}. These calculations are economical compared to the self-consistent methods that suffer from the boundary problems \cite{Avancini_2008}, and hence are applied in a wide range of inner crust calculations.

We  calculate the equilibrium composition of the outer crust using the most recent experimental atomic mass evaluations AM2020 \cite{Huang_2021} in supplement with the theoretical calculations of Hartree-Fock-Bogoliubov (HFB) \cite{hfb2426, hfb14} and finite-range liquid-drop model (FRDM) \cite{MOLLER20161}. Various inner crust properties, such as equilibrium composition, crust-core transition properties, pasta phase appearance, pasta mass, thickness, and frequency of QPOs in context to the soft gamma repeaters (SGRs), etc., are calculated for a magnetized neutron star. These properties play a central role in explaining various mechanisms of magnetar and pulsar activities which include transport of magnetic field lines (hall drift) \cite{Konstantinos_2016, Perna_2011}, sudden fractures in the crust due to accumulating crustal stress \cite{Beloborodov_2014}, neutron superfluidity that causes glitches in pulsars \cite{anderson1975pulsar}, transient heating of crust \cite{Li_2015}, etc. Furthermore, we also investigate the role of EoS on the crustal properties of neutron stars in a magnetized environment. 

The paper's organization is as follows: In Sec \ref{formulation}, we describe the effect of magnetic field on the EoS employing the E-RMF framework. The outer and inner crust formulation is described in brief. The results are discussed in Sec. \ref{results}, where the outer crust is discussed in Sec. \ref{oc}, the inner crust and associate properties within CLDM formalism in Sec. \ref{ic}, and the unified EoS in Sec. \ref{unified}. Finally, we summarize our results in Sec. \ref{conclusion}.
\section{Formalism}
\label{formulation}
\subsection{E-RMF in the presence of magnetic field}
The effective Lagrangian in the effective relativistic mean field model (E-RMF), which include  $\sigma$, $\omega$, $\rho $, $\delta$, and photon in association with the baryons  can be written as  \cite{Patra_2002, MULLER_1996, Wang_2000, DelPairing_2001, Kumar_2020, Das_2021},
\begin{widetext}
\begin{eqnarray}
\label{rmftlagrangian}
\mathcal{E}(r)&=&\psi^{\dagger}(r)\qty{i\alpha\cdot\grad+\beta[M-\Phi(r)-\tau_3D(r)]+W(r)+\frac{1}{2}\tau_3R(r)+\frac{1+\tau_3}{2} A(r)-\frac{i\beta \alpha }{2M}\qty(f_\omega \grad W(r)+\frac{1}{2}f_\rho \tau_3 \grad R(r))}\psi(r) \nonumber \\
&+& \qty(\frac{1}{2}+\frac{k_3\Phi(r)}{3!M}+\frac{k_4}{4!}\frac{\Phi^2(r)}{M^2})\frac{m^2_s}{g^2_s}\Phi(r)^2+\frac{1}{2g^2_s}\qty\Big(1+\alpha_1\frac{\Phi(r)}{M})(\grad \Phi(r))^2-\frac{1}{2g^2_\omega}\qty\Big(1+\alpha_2\frac{\Phi(r)}{M})(\grad W(r))^2 \nonumber\\
&-&\frac{1}{2}\qty\Big(1+\eta_1\frac{\Phi(r)}{M}+\frac{\eta_2}{2}\frac{\Phi^2(r)}{M^2})\frac{m^2_\omega}{g^2_\omega}W^2(r)-\frac{1}{2e^2}(\grad A^2(r))^2 -\frac{1}{2g^2_\rho}(\grad R(r))^2
-\frac{1}{2}\qty\Big(1+\eta_\rho\frac{\Phi(r)}{M})\frac{m^2_\rho}{g^2_\rho}R^2(r)\nonumber \\
&-&\frac{\zeta_0}{4!}\frac{1}{g^2_\omega}W(r)^4-\Lambda_\omega(R^2(r)W^2(r))
+\frac{1}{2g^2_\delta}(\grad D(r))^2
+\frac{1}{2}\frac{m^2_\delta}{g^2_\delta}(D(r))^2.
\end{eqnarray} 
\end{widetext}
Here $\Phi(r)$, $W(r), R(r), D(r)$ and $A(r)$ are the fields corresponding to $\sigma$, $\omega$, $\rho$ and $\delta $ mesons and photon, respectively. The $g_s$, $g_{\omega}$, $g_{\rho}$, $g_{\delta}$ and $\frac{e^2}{4\pi }$ are the corresponding coupling constants and $m_s$, $m_{\omega}$, $m_{\rho}$ and $m_{\delta}$ are the corresponding masses. The zeroth component $T_{00}= H$ and the third component $T_{ii}$ of energy-momentum tensor \cite{NKGb_1997}
\begin{equation}
\label{set}
T_{\mu\nu}=\partial^\nu\phi(x)\frac{\partial\mathcal{E}}{\partial(\partial_\mu \phi(x))}-\eta^{\nu\mu}\mathcal{E},
\end{equation}
yields the energy and pressure density, respectively.

Various couplings in the E-RMF Lagrangian have their own importance and make the model flexible to accommodate various phenomena associated with nuclear matter. The $\zeta_0$ (self-coupling of isoscalar-vector $\omega$ meson ($W (r)^4$)) and the self-coupling of $\sigma$ meson ($k_3$, $k_4$) help soften the equation of state. On the other hand, $\Lambda_\omega$ (quartic-order cross-coupling of $\rho$ and $\omega$  meson ($R(r)^2W (r)^2$)) plays a crucial role in governing the density dependence of symmetry energy and help in the better agreement of neutron skin thickness data. $\Lambda_\omega$ also provide us with the flexibility of fitting the spherical nuclei without hurting the ability to vary the neutron skin thickness of $^{208}$ Pb to a wide range \cite{Tapas_2005}. $\eta_1, \eta_2, \eta_\rho$   couplings (cross-couplings of $\sigma-\omega$ and $\sigma-\rho$ mesons) influence the surface properties of finite nuclei and the $\delta$ meson softens the symmetry energy at subsaturation densities while stiffens the EoS at high densities \cite{KUBIS1997191, Singh_2014}. For more details on the E-RMF formalism and EoS, one can see \cite{Kumar_2018, Patra_2002, Aggarwal_2010}. Here we briefly describe the effect of magnetic field on the EoS, based on \cite{Broderick_2000, Strickland_2012}.

The energy spectrum of the proton, which gets modified due to the Landau level, is written as \cite{Broderick_2000, Strickland_2012}
\begin{equation}
    E_p=\sqrt{k_z^2+\ols{M}_{n,\sigma_z}^{p^2}}+W-R/2,
\end{equation}
and for charged leptons (electron and muon) as
\begin{equation}
    E_{e,\mu}=\sqrt{k_z^2+\ols{M}_{n,\sigma_z}^{{e,\mu}^2}},
\end{equation}
where 
\begin{align}
        \ols{M}_{n,\sigma_z}^{(p)^2}=M_{(p)}^{*^2}+2\Big(n+\frac{1}{2}-\frac{1}{2}\frac{q}{|q|}\sigma_z \Big)|q|B.\\
        \ols{M}_{n,\sigma_z}^{(e,\mu)^2}=M_{(e,\mu)}^{2}+2\Big(n+\frac{1}{2}-\frac{1}{2}\frac{q}{|q|}\sigma_z \Big)|q|B.
\end{align}
Here, $\sigma_z$ is the spin along the axis of magnetic field $B$, $n$ is the principal quantum number, and $k_z$ is the momentum along the direction of the magnetic field. $M^*$ is the effective mass for the proton. The neutron spectrum is similar to the Dirac particle and takes form. 
\begin{equation}
    E_n=\sqrt{k^2+M_n^{*^2}}+W+R/2.
\end{equation}
The number and energy density at zero temperature and in the presence of a magnetic field is given by \cite{Broderick_2000}
\begin{equation}
\label{eq:density}
    \rho_{i=e,\mu,p}=\frac{|q|B}{2 \pi^2} \sum_{\sigma_z} \sum_{n=0}^{n_{max}}k_{f,n,\sigma_z}^{i},
\end{equation}
\begin{eqnarray}
\label{eq:energy}
    E_{i=e,\mu,p}&=&\frac{|q|B}{4 \pi^2} \sum_{\sigma_z} \sum_{n=0}^{n_{max}} \nonumber\\
    &\cross& \Big[E_f^{i}k_{f,n,\sigma_z}^{i}     + \ols{M}_{n,\sigma_z}^{{i^2}} \ln \Big(\Big|\frac{E_f^i+k_{f,n,\sigma_z}^i}{\ols{M}_{n,\sigma_z}^{{i^2}}}\Big|\Big)\Big].
\end{eqnarray}
respectively. In above equations, $k_{f,n,\sigma_z}^i$ is defined by
\begin{equation}
\label{eq:fermi_momentum}
    k_{f,n,\sigma_z}^{i^2}=E_f^{i^2}-\ols{M}_{n,\sigma_z}^{{i^2}},
\end{equation}
where the Fermi energies are fixed by the respective chemical potentials given by
\begin{align}
    E_f^{l=e,\mu}=\mu_{\mu,e},\\
    E_f^{b=p,n}=\mu_b-W \pm R/2.
\end{align}
In Eq. (\ref{eq:density}) and (\ref{eq:energy}), the $n_{\rm max}$ is the integer for which the Fermi momentum remains positive in Eq. (\ref{eq:fermi_momentum}) and is written as
\begin{align}
    &n_{max}=\Bigg[\frac{E_f^{i^2}-M^{*^2}}{2|q|B}\Bigg], \; {\rm proton}\\ \nonumber
    &n_{max}=\Bigg[\frac{E_f^{i^2}-M^2}{2|q|B}\Bigg], \; \; {\rm electron} \; \& \; {\rm muon} \, .
\end{align}
Here $[x]$ represents the greatest integer less than or equal to $x$. The scalar density for the protons is further determined as 
\begin{equation}
    \rho_p^s=\frac{|q|BM^*}{2\pi^2}\sum_{\sigma_z} \sum_{n=0}^{n_{max}}\ln \Big(\Big|\frac{E_f^i+k_{f,n,\sigma_z}^i}{\ols{M}_{n,\sigma_z}^{{i^2}}}\Big|\Big).
\end{equation}
The number, scalar, and energy density for the neutrons are similar to the field-free case and can be found in \cite{Kumar_2018, Patra_2002} and reference therein. The total energy density is the sum of matter-energy density and the contribution from the electromagnetic field, $\frac{B^2}{8 \pi}$. Finally, the pressure can be written, keeping the thermodynamic consistency as
\begin{equation}
\label{eq:press}
P= \sum_{i=n,p}\mu_i\rho_i-E \, .   
\end{equation}
It is often convenient to express the magnetic field strength in the critical magnetic field of an electron ($B_c$) as
\begin{equation}
    B^*=\frac{B}{B_c}, 
\end{equation}
where $B_c \sim 4.414 \cross 10^{13}$ G. A magnetic field strength is said to be strongly quantizing if only the lowest Landau level is filled i.e. $\nu=\Big(n+\frac{1}{2}-\frac{1}{2}\frac{q}{|q|}\sigma_z \Big)=0$ \cite{Haenel_2007}.
\subsection{Outer and inner crust}
The outer crust is assumed to be stratified into various layers, which consist of a single nuclear species (single-nucleus approximation \cite{Shen_2011}) surrounded by relativistic degenerate electron gas.  The nuclei make up a perfect crystal arranged on a body-centered cubic (bcc) lattice. To determine the properties of the outer crust in the presence of magnetic field $B$, we minimize the Gibbs free energy given by \cite{Carreau_2020}
\begin{equation}
\label{eq:gibbsminimization}
    G(A,Z,P)=\frac{M(A,Z)}{A}+\frac{4}{3}\frac{E_L}{A}+\frac{1}{2}\frac{E_{zp}}{A}+\frac{Z}{A}\mu_e,
\end{equation}
at a fixed pressure, where $\mu_e$ is the electron chemical potential, $E_L$ and $E_{zp}$ are the static-lattice and zero-point energy \cite{Parmar_2022_1,Carreau_2020}. The nuclear mass, $M(A, Z)$, is obtained from either the experimental atomic mass evaluations \cite{Huang_2021} or theoretical calculations such as the HFB framework. The condition $\mu_n=M_n$, where $\mu_n$ is the chemical potential of neutrons, makes the outer crust boundary where neutrons start to drip out of the nuclei. 

The inner crust comprises clusters surrounded by neutron gas and ultra-relativistic electron gas.  Using the CLDM based on the Wigner-Seitz (WS) cell approximation, the energy of the cluster is written as \cite{Parmar_2022_1, Carreau_2019, Newton_2013, Dinh_2021}
\begin{align}
    E(r_c, y_p,\rho, \rho_n)&=f(u)\left[E_{\rm bulk}(\rho_b, y_p)\right]
    \nonumber\\
    &
    +E_{\rm bulk}(\rho_g,0)\left[1-f(u)\right]
    \nonumber\\
    &
    +E_{\rm surf}+E_{\rm curv}+E_{\rm coul}+E_e.
    \label{eq:energy_ic}
\end{align}
Here $r_c$ is the radius (half-width in the case of planar geometry) of the WS cell, $y_p$ is the proton fraction, and $\rho$ and $\rho_n$ are the baryon density of charged nuclear component, and density of neutron gas, respectively. The cluster is characterised by density $\rho_i$ and volume fraction $u$ as \cite{Newton_2012, Dinh_2021}
\begin{equation}
    u=\begin{cases}
       (\rho-\rho_g)/(\rho_i-\rho_g) \, \, \, \text{for clusters},\\ (\rho_i-\rho)/(\rho_i-\rho_g) \, \, \, \, \text{for holes}.
      \end{cases}
\end{equation}
The function $f(u)$ is defined as 
\begin{equation}
    f(u) = \begin{cases}
           u  \, \, \, \, \,  \text{for clusters}, \\
           1-u \, \, \, \text{for holes.}
          \end{cases}
\end{equation}

We consider the three canonical geometries, namely spherical, cylindrical, and planar, defined by a dimensionality parameter $d = 3, 2, 1,$ respectively. The surface and curvature energies are written as \cite{Newton_2013, Dinh_2021},
\begin{equation}
    E_{\rm surf}+E_{\rm curv}=\frac{u d}{r_N}\left( \sigma_s +\frac{(d-1)\sigma_c}{r_N}\right),
\end{equation}
where $r_N$ is the radius/half-width of the cluster/hole and $\sigma_s$ and $\sigma_c$ are the dimension-independent surface and curvature tension  based on the TF calculations and are defined as \cite{Ravenhall_1983} 
\begin{equation}
\label{surf}
    \sigma_s=\sigma_0\frac{2^{p+1} + b_s}{y_p^{-p} +b_s+(1-y_p)^{-p}},
\end{equation}
\begin{equation}
\label{curv}
    \sigma_c=\alpha \, \sigma_s\frac{\sigma_{0,c}}{\sigma_0}\left(\beta-y_p\right).
\end{equation}
Here the parameters ($\sigma_0$, $\sigma_c$, $b_s$, $\alpha$, $\beta$, $p$) are optimized for a given equation of state on the atomic mass evaluation 2020 data \cite{Huang_2021}. The Coulomb energy reads as \cite{Dinh_2021}
\begin{equation}
    E_{\rm coul}=2\pi (e\,y_p\,n_i\,r_N)^2\,u\,\eta_d(u),
\end{equation}
where $e$ is the elementary charge and $\eta_d(u)$ is associated with the pasta structures. The detailed formalism for the CLDM approximation is given in \cite{Parmar_2022_1, Dinh_2021, Carreau_2019}.

For a given baryon density, the equilibrium composition of a WS cell is obtained by minimizing the energy per unit volume using the variational method where the auxiliary function to be minimized reads as \cite{Parmar_2022_1, Carreau_2019}
\begin{equation}
    \mathcal{F}=\frac{E_{\rm WS}}{V_{\rm WS}}-\mu_b\rho.
\end{equation}
Here, $\mu_b$ is the baryonic chemical potential. This results in a set of four differential equations corresponding to mechanical, dynamical, $\beta$-equilibrium, and the nuclear virial theorem \cite{Carreau_2020, Carreau_2019}. The viral relation is used to solve the value of $r_N$ numerically. To obtain the most stable pasta structure at a given baryon density, we first calculate the composition of a spherical nucleus. Then keeping this composition fixed, we calculate the radius or half-width of five different pasta structures, namely, sphere, rod, plate, tube, and bubble. The equilibrium phase is then the one that minimizes the total energy of the system. 

\subsection{Neutron star observables}
\label{form:NS_observable}
For a static star, the macroscopic properties such as $M$ and $R$ of the neutron star can be found  by solving the following Tolmann-Oppenheimer-Volkoff (TOV) equations as \cite{TOV1, TOV2}
\begin{eqnarray}
\frac{dP(r)}{dr}= - \frac{[P(r)+{\cal{E}}(r)][m(r)+4\pi r^3 P(r)]}{r[r-2m(r)]},
\label{eq:pr}
\end{eqnarray}
and 
\begin{eqnarray}
\frac{dm(r)}{dr}=4\pi r^2 {\cal{E}}(r).
\label{eq:mr}
\end{eqnarray}
The $M$ and $R$ of the star can be calculated with boundary conditions $r=0, P = P_c$ and $r=R, P = P_0$ at certain central density.

The neutron star's moment of inertia (MI) is calculated in the Refs. \cite{Stergioulas_2003,Jha_2008,Sharma_2009,Friedmanstergioulas_2013,Paschalidis_2017,Quddus_2019,Koliogiannis_2020}. The expression of $I$ of uniformly rotating neutron star with angular frequency $\omega$ is given as \cite{Hartle_1967,Lattimer_2000,Worley_2008}
\begin{equation}
I \approx \frac{8\pi}{3}\int_{0}^{R}\ dr \ ({\cal E}+P)\  e^{-\phi(r)}\Big[1-\frac{2m(r)}{r}\Big]^{-1}\frac{\Bar{\omega}}{\Omega}\ r^4,
\label{eq:moi}
\end{equation}
where $\Bar{\omega}$ is the dragging angular velocity for a uniformly rotating star. The $\Bar{\omega}$ satisfying the boundary conditions are 
\begin{equation}
\Bar{\omega}(r=R)=1-\frac{2I}{R^3},\qquad \frac{d\Bar{\omega}}{dr}\Big|_{r=0}=0 .
\label{eq:omegabar}
\end{equation}
To calculate the accurate core/crust thickness or mass, one needs to integrate the TOV Eqs. \ref{eq:pr} and \ref{eq:mr} from $R=0$ to $R=R_{\rm core}$, which depends on pressure as $P(R=R_{\rm core})=P_t$. We calculate the crustal MI by using the Eq. (\ref{eq:moi}) from the transition radius ($R_c$) to the surface of the star ($R$), which is given by \cite{Fattoyev_2010, Basu_2018} 
\begin{equation}
I_{\rm crust} \approx \frac{8\pi}{3}\int_{R_c}^{R}\ dr\ ({\cal E}+P)\  e^{-\phi(r)}\Big[1-\frac{2m(r)}{r}\Big]^{-1}\frac{\Bar{\omega}}{\Omega}\ r^4.
\label{eq:moic}
\end{equation}
The second Love number ($k_2$) and its corresponding dimensionless tidal deformability ($\Lambda$) are related with a relation $\Lambda = (2/3)k_2 C^5$, where $C$ is the compactness of the star \cite{DasBig_2020}. The details regarding the Love number and tidal deformability can be found in Refs. \cite{Hinderer_2008, Kumar_2018, DasBig_2020}. 
\section{Results and Discussion}
\label{results}
In this section, we discuss the effect of the magnetic field on the outer and inner crust of a neutron star and its implications on crustal properties. The only input parameters in calculating the outer crust composition are the mass of the nuclei and magnetized electron gas EoS. In this work, the experimental masses are taken from the most recent mass evaluations AME2020 \cite{Huang_2021} whenever available, along with the data of $^{82}$Zn \cite{wolf2013}, $^{77-79}$Cu \cite{welker2017}, $^{151-157}$Yb \cite{Beck_2021}. For unknown masses, we use the theoretical microscopic calculations of HFB24, and HFB26 \cite{hfb2426}, which are highly sophisticated mass models based on accurately calibrated Brussels-Montreal functional for unconventional regimes. In addition, we also use the most recent FRDM \cite{MOLLER20161} results for comparison.
\begin{figure*}
  \centering
\subfloat[]{%
  \label{fig:fig1sub1}
  \includegraphics[width=0.49\linewidth]{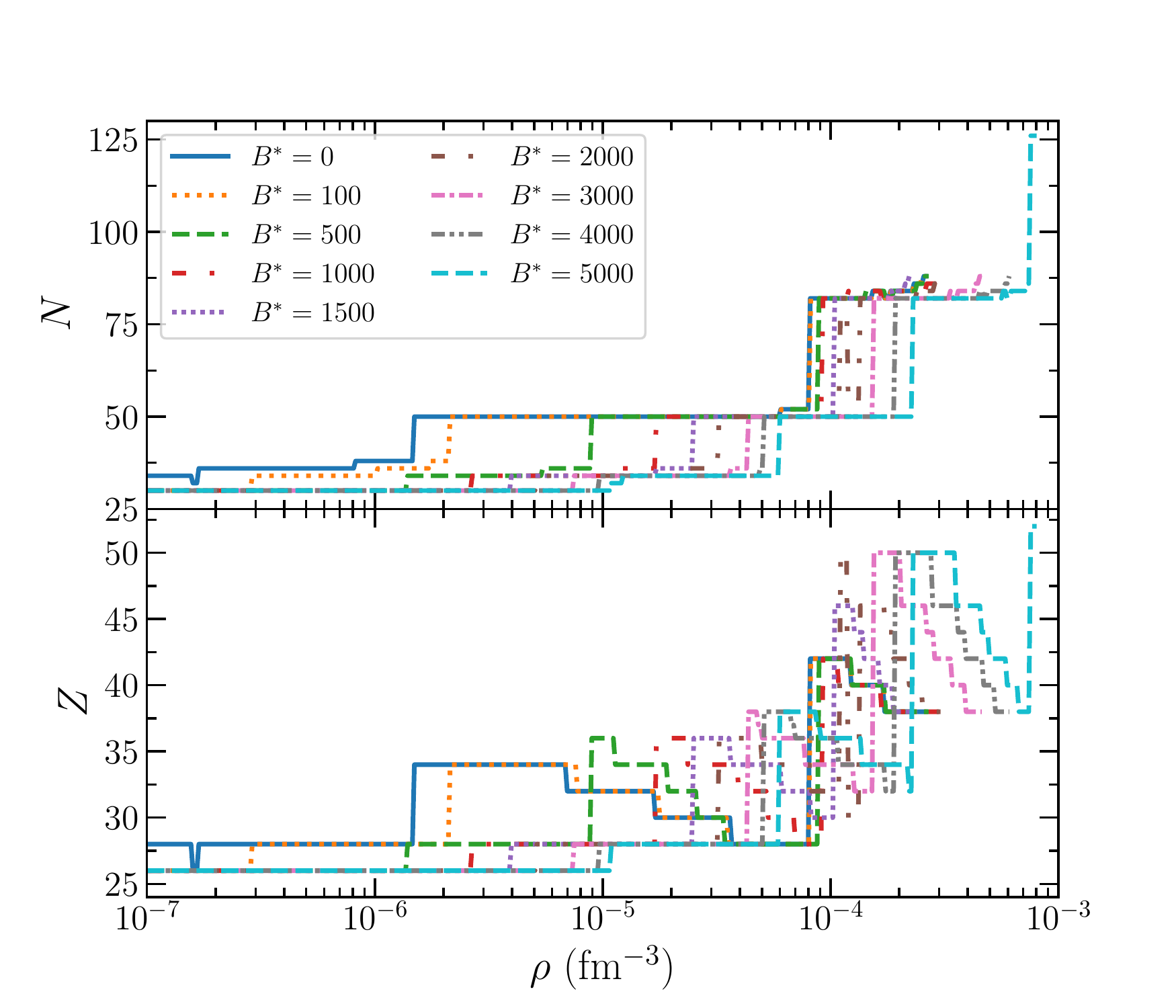}%
}\hfill
\subfloat[]{%
  \label{fig:fig1sub2}
  \includegraphics[width=0.49\linewidth]{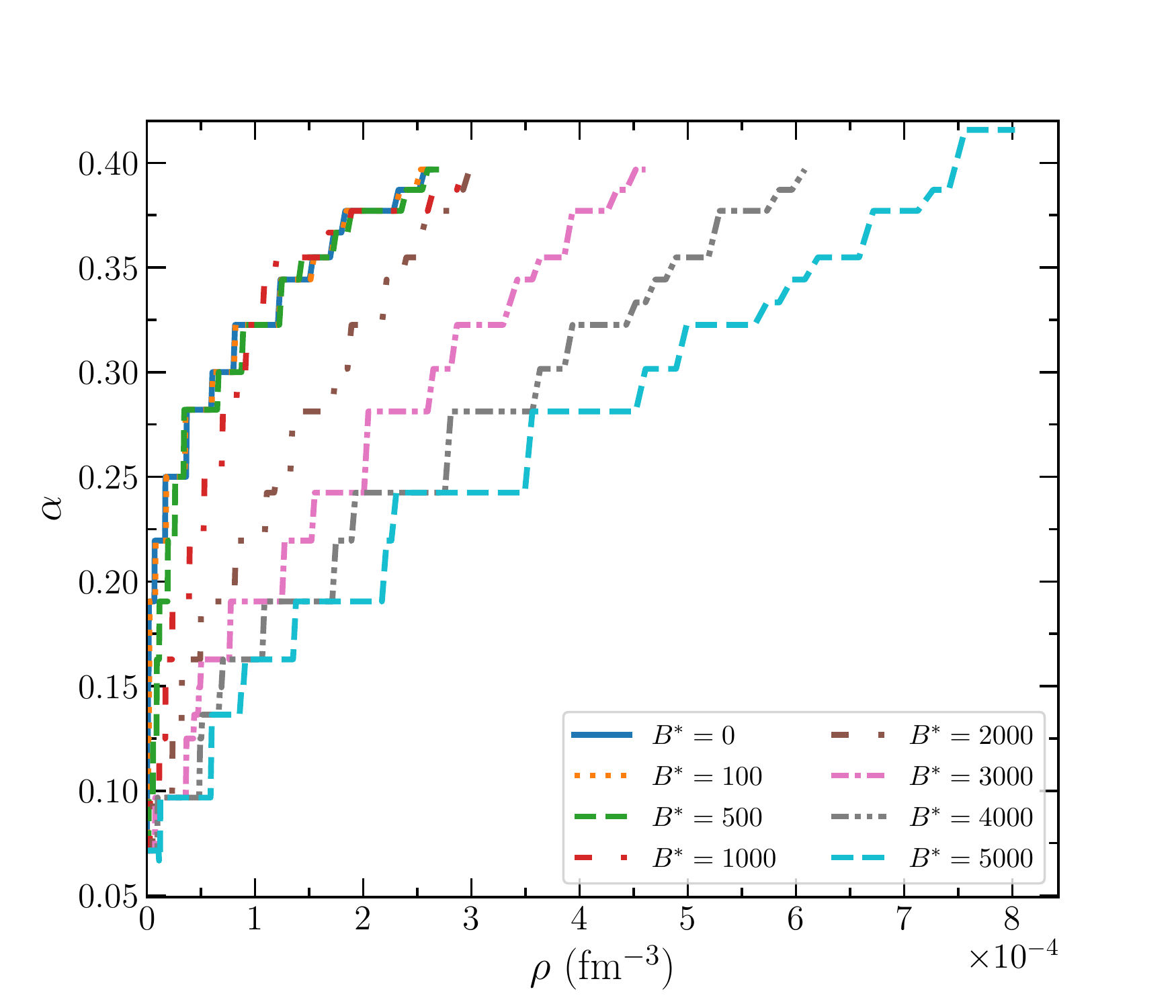}%
} 
\caption{(a) The neutron number ($N$) and charge/proton number ($Z$) as a function of density in the outer crust of magnetized neutron star for various values of magnetic field strength $B^*$. The unknown masses are taken from the HFB26 mass model. (b) The variation of $\alpha=\frac{N-Z}{N+Z}$ as a function of density. }
\end{figure*}

The inner crust structure is determined using six E-RMF parameter sets with varying saturation properties, namely: G3 \cite{Kumar_2017}, IOPB-I \cite{Kumar_2018}, FSUGarnet \cite{Chen_2015}, IUFSU \cite{Fattoyev_2010}, IUFSU$^*$ \cite{Fattoyev_2010}  and SINPB \cite{Mondal_2016}.  In Refs. \cite{Parmar_2022_1, Parmar_2022_2}, we used these parameter sets to  study a neutron star's crustal properties without considering the magnetic field's effect. Here we assess the magnetic effects on the EoS and intend to investigate the corresponding  effect on neutron star properties. We also provide the unified EoS at a given magnetic field, considering the same strength throughout the neutron star's interior. We restrict the strength of the magnetic field under $B^* \le 5000$ to satisfy the assumptions made in \cite{Patra_2020} to calculate the neutron star structure using spherically symmetric treatment of the NS structure.
\subsection{Outer crust}
\label{oc}
\begin{figure*}
    \centering
    \includegraphics[scale=0.45]{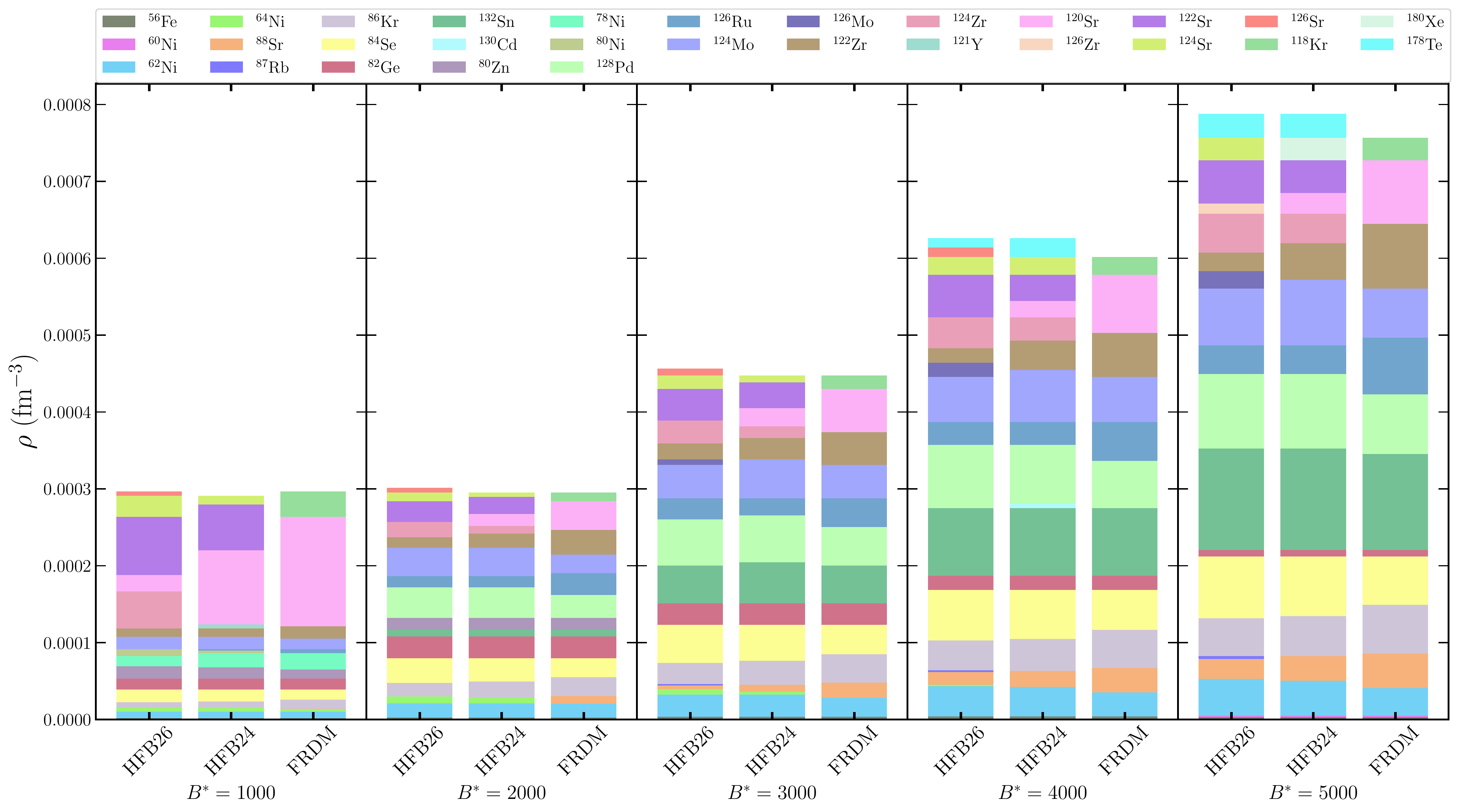}
    \caption{The sequence of nuclei in the outer crust of neutron star from low to high density at various magnetic field values. The most recent experimental atomic mass evaluations AME2020 \cite{Huang_2021} are taken whenever available and microscopic calculations of HFB24 and HFB26 \cite{hfb2426} along with the FRDM(2012) mass tables \cite{MOLLER20161} are used for the unknown mass.}
    \label{fig:nuc_seq}
\end{figure*}
\begin{table*}
\caption{The surface density of the outer crust ($\rho_{\rm surf}$) and neutron drip ($\rho_{\rm drip}$)  for the magnetic fields and mass models considered.}
\label{tab:oc_dens}
\renewcommand{\arraystretch}{1}
\resizebox{\textwidth}{!}{%
\begin{tabular}{llllllllllllllll}
\hline
\hline
$B^*$ & \multicolumn{3}{c}{1000} & \multicolumn{3}{c}{2000} & \multicolumn{3}{c}{3000} & \multicolumn{3}{c}{4000} & \multicolumn{3}{c}{5000} \\
\hline
$\rho$ (fm$^{-3}$)  & HFB26   & HFB24  & FRDM  & HFB26   & HFB24  & FRDM  & HFB26   & HFB24  & FRDM  & HFB26   & HFB24  & FRDM  & HFB26   & HFB24  & FRDM  \\
   \hline
 $\rho_{\rm surf}$ $\cross$ ($10^{-6}$) &1.0554  &1.0544 &1.0544 & 2.4708 & 2.4708 & 2.4708  &4.0537  & 4.0537  &4.0537 & 5.7897 &5.7897 &5.7897 &7.5783  &7.5783  & 7.5783  \\
 $\rho_{\rm drip}$ $\cross$ ($10^{-4}$)& 2.9791  &2.9207  & 2.9791 &3.0387& 2.9791  &   2.9791 &4.6057 &4.5154  &4.5154  & 6.3226 &6.3226  &  6.0771 & 7.9545  & 7.9545 &7.6456  \\
\hline \hline
\end{tabular}%
}
\end{table*}

The equilibrium composition of the outer crust of a nonaccreting magnetized neutron star is determined by minimizing the Gibbs free energy given in Eq. \ref{eq:gibbsminimization} at a fixed pressure. In principle,  the nuclear masses should modify in the presence of the magnetic field \cite{Arteaga_2011, Stein_2016, Basilico_2015}, which might affect the outer crust composition. However, a comprehensive mass table in the presence of a magnetic field is not yet available.  Moreover, Refs. \cite{Arteaga_2011, Stein_2016, Basilico_2015}  suggest that a field strength $> 10^{17} G$ is required to alter the nuclear ground state significantly, whereas the highest observed field strength at the surface of the magnetar is $20 \cross 10^{14} G$ for SGR 1806-20 \cite{Olausen_2014} among  26 currently known magnetars \footnote{\url{https://www.physics.mcgill.ca/~pulsar/magnetar/main.html}}. Therefore we use the nuclear masses for the field-free case in this study, keeping the strength of magnetic field $\sim 10^{17}$ G.

In Fig. \ref{fig:fig1sub1}, we show the equilibrium composition of the outer crust of a magnetized neutron star as a function of density at various strengths of the magnetic field $B^*=B/B_c$. The nuclear masses are  taken from  experimental AME2020 \cite{Huang_2021}, and HFB26 \cite{hfb2426} table. The Outer crust is stratified into various layers. For a weak magnetic field ($B^*\sim 10$), which is relevant for the ``pulsar" \cite{becker2009neutron}, the composition remains similar to the field-free case. For the field strength $B^*>500$, the sequence of nuclei and the density at which they occur change significantly. The $Z=26$ and $N=30$ ($^{56}$Fe) plateau keeps extending with the increasing magnetic field and extends up to 1.0706E-5 fm$^{-3}$ for the field strength $B^*=5000$ as compared to the 4.9729E-9 fm$^{-3}$ for the field free case. The density at which the $N=50$ plateau appears ( characteristic of the outer crust at $B^*=0$), increases monotonically with increasing field strength. This means that the nuclei become more and more symmetric with increasing magnetic fields at the same pressure. This is clear from Fig. \ref{fig:fig1sub2} where we plot the asymmetry parameter $\alpha=\frac{N-Z}{N+Z}$ as a function of density. The isospin asymmetry increases almost linearly at higher magnetic field whereas an exponential behavior is observed for the field-free case. However, the maximum $\alpha$ does not exceed $\sim 0.4$ for field strength as high as $B^*=5000$. This behavior of the outer crust composition can be attributed to the EoS of the electron gas. With increasing field strength, the chemical potential or Fermi energy of the electrons decreases, which enforces a delay in the pressure where new nuclear species stat appearing. 

The qualitative observations in Fig. \ref{fig:fig1sub1} and \ref{fig:fig1sub2} are also supported by the nuclear mass model HFB24 \cite{hfb2426}, and FRDM \cite{MOLLER20161}. The detailed composition and equation of state for these mass models, along with the unified magnetized equation of states, are provided in the GitHub link \footnote{\url{https://github.com/hcdas/Unified_mf_eos}}. As the deeper portion of the outer crust is determined using the mass excess from the theoretical mass models, there exists a model dependency of the sequence of the nuclear species \cite{Parmar_2022_1}. To investigate the model dependency in the presence of the magnetic field, we show the sequence of nuclear species in Figure \ref{fig:nuc_seq} along with the density of the surface ($\rho_{\rm surf}$), and neutron drip density ($\rho_{\rm drip}$) in Table \ref{tab:oc_dens}  for the HFB26, HFB24, and FRDM mass models at various magnetic field strengths. 

\begin{table}
\centering
\caption{Symmetry energy ($J$) and slope parameter ($L$) coefficient for the HFB26, HFB24, and FRDM mass models.}
\label{tab:mass_model_sym}
\scalebox{1.0}{
\begin{tabular}{llll}
\hline
\hline
& HFB24 \cite{Pearson_2018}  & HFB26 \cite{Pearson_2018}  & FRDM \cite{Hiroyuki_frdmsym}  \\
\hline    
$J$ (MeV) & \hfill 30.0   & \hfill 30.0   &  \hfill 32.3  \\
$L$ (MeV) & \hfill 37.5   & \hfill 46.4   &  \hfill 53.5  \\
\hline

\end{tabular}%
}
\end{table}

All the mass models predict approximately the same sequence of nuclei for a given magnetic field strength, except for the appearance/disappearance of certain nuclei. This shows that the outer crust of a neutron star is dominantly dependent on the structural effects (magic number of neutrons at $N=50, 82$) of the nucleus rather than the nuclear matter properties of the EoS with which their masses have been determined. We show the symmetry energy ($J$) and slope parameter ($L$) of the mass models in Table \ref{tab:mass_model_sym} for reference. The outer crust preserves the $N=50 \, {\rm and} \, 80$ plateau for the magnetic field as high as $B^*=5000$. When comparing different mass models, we observe that the FRDM estimates a constant $N=82$ nuclei while the HFB26 and HFB24 mass models deviate from it at higher density. The FRDM mass model estimates the presence of $^{118}$Kr at the neutron drip density for all the magnetic field strengths, while the HFB26 and HFB24 ( having similar symmetry energy), estimate the different isotopes of Sr for $B^*$ up to $3000$ and $^{178}$Te at higher magnetic field strength.

Furthermore, as we increase the magnetic field strength, $^{132}$Sn appears in place of $^{80}$Zn. There is also a possibility of the existence of  $^{130}$Cd at higher magnetic field strength, although for a brief span. The odd number nucleus $^{121}$Y, observed for the field-free case using the HFB24 mass model, disappears at higher magnetic field strength.  For a particular value of the magnetic field, it is observed that the transition of one nucleus to another happens at a lower density for FRDM as compared to the HFB24, followed by HFB26 in the regions closer to neutron drip density (where the role of the theoretical mass model comes into play). This trend can be attributed to their decreasing slope parameter $L$.

With an increase in magnetic field strength, the surface density of the outer crust of the neutron star increases as high as seven times when compared with the $B^*=1000$ and $B^*=5000$ case due to the magnetic condensation. This surface density is determined b on the basis of experimental mass of $^{56}$Fe and hence is the same for all the mass models. The neutron drip density increases exponentially with increasing magnetic field and becomes as high as three times compared to the $B^*=1000$. However, the neutron drip density exhibits an oscillatory quantum nature at a lower magnetic field arising from the Landau quantization of electrons resulting in the increase or decrease of neutron drip density. Similar behavior was  demonstrated in \cite{Chamel_2015_landauquant}, which suggests the model independency of these oscillations, which occur for approximately $B^*<1300$. For higher magnetic field strengths, the electron EoS becomes strongly quantizing in the whole outer crust. A slight variation in the neutron drip density for various mass models at a given magnetic field strength is the consequence of the drip nuclei which these models predict. More neutron-rich (less bound) nuclei can sustain at lower pressure for a higher magnetic field. The asymmetry at the neutron drip density thus keeps increasing with increase in magnetic field strength. 
\begin{figure}
    \centering
    \includegraphics[scale=0.5]{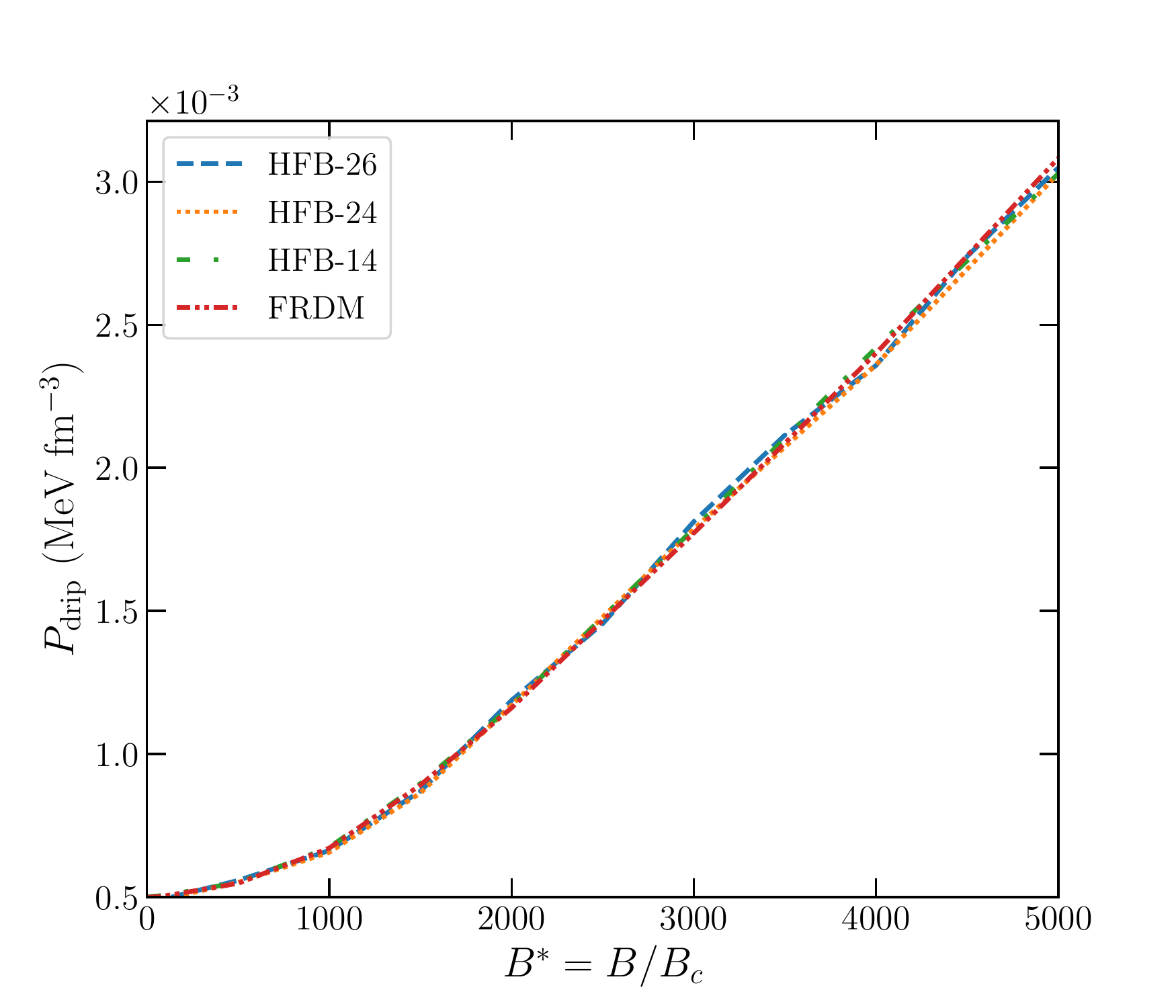}
    \caption{Transition pressure ($P_{\rm drip}$) at the neutron drip density in the crust of neutron star as a function of magnetic field strength.}
    \label{fig:pdrip_oc}
\end{figure}

The transition pressure at the neutron drip point is more important than the transition density, as the former plays a direct role in calculating the mass of the crust \cite{Pearson_2018, Parmar_2022_1}. We plot the transition pressure for the various mass models: HFB14, HFB24, HFB26, and FRDM in Fig. \ref{fig:pdrip_oc} as a function of magnetic field strength. It increases linearly for a magnetic field greater than $\sim 1300$, which is the strongly quantizing regime (only the lowest Landau level $\nu=0$ is filled). Transition pressure increases almost six times for $B^*=5000$ as compared to the field-free case. A higher neutron drip transition pressure implies that the crustal mass of the outer crust will be larger for greater strength of the magnetic field. Furthermore, the neutron drip pressure as a function of magnetic field strength seems to be model-independent. Chamel {\it et al.} \cite{Chamel_2012} have determined analytic expression for the outer crust transition pressure ($P_{\rm drip}$) in the strongly quantizing regime (B$^* >$ 1300). Based on the evident model independency of the $P_{\rm drip}$ in Fig. \ref{fig:pdrip_oc}, the $P_{\rm drip}$ as a function of the magnetic field in both quantizing and non-quantizing regime can be written as

\begin{align}
    P_{\rm drip}=&-1.81110^{-14}\cross B^{*3}+1.84910^{-10}\cross  B^{*2} \nonumber \\ 
    &+3.57810^{-8}\cross B^*+0.00049,    
\end{align}
where the coefficients have relevant dimensions in the form of MeV fm$^{-3}$.
\begin{figure}
    \centering
    \includegraphics[scale=0.5]{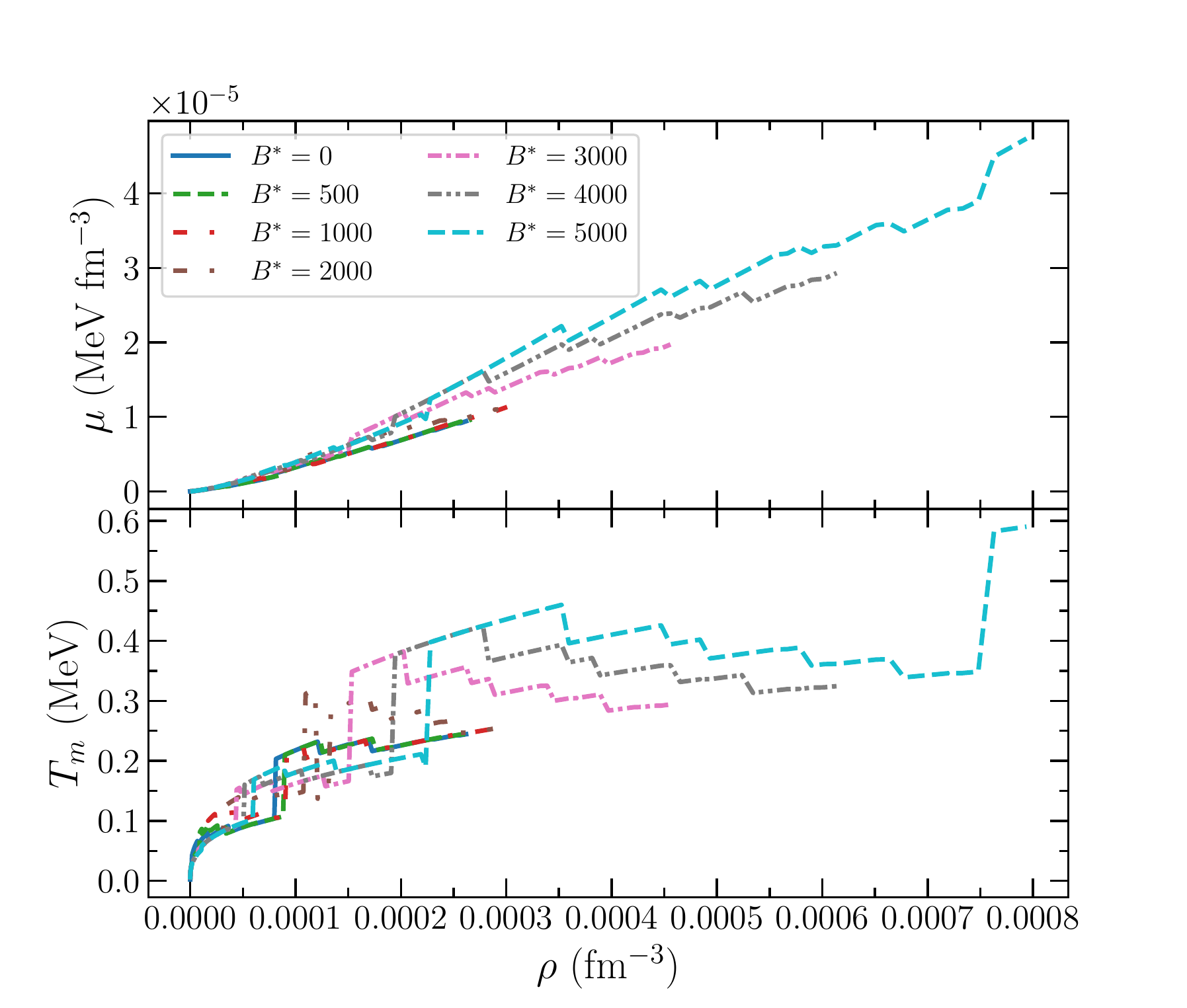}
    \caption {The variation of shear modulus $\mu$ and crystallization temperature $T_m$ in the outer crust of a neutron star at various magnetic field strengths. The theoretical mass model is HFB26}
    \label{fig:oc_tmsm}
\end{figure}

Other essential aspects of the outer crust of a neutron star are its mechanical responses and melting temperature, which plays a prime role in describing properties such as crust failure \cite{Beloborodov_2014} and dynamics of the crust \cite{Newton_2012}. The mechanical properties are determined using its shear stress which for a cold neutron star can be written  following Monte Carlo simulation \cite{Chugnov_2010} as
\begin{equation}
    \mu=0.1106\Big(\frac{4 \pi}{3}\Big)^{1/3}A^{-4/3}\rho_i^{4/3}(1-X_n)^{4/3}(Ze)^2,
\end{equation}
where $\rho_i$ is the density of nuclei, and $X_n$ is the fraction of neutrons not confined to the nuclei. The melting or crystalline temperature, which defines the temperature at which the crystalline lattice converts to the gas of ions, is written in the one-component plasma (OCP) approximation as \cite{Newton_2012, Fatima_2020}
\begin{equation}
    T_m= \frac{Z^2e^2}{k_b \Gamma_m}\Big(\frac{4 \pi}{3} \frac{\rho_i}{A} \Big)^{1/3}.
\end{equation}
Here, $\Gamma_m=175$ is the Coulomb coupling parameter at melting. We show the variation of shear modulus ($\mu$) and melting temperature ($T_m$) in the outer crust in Fig. \ref{fig:oc_tmsm} at various magnetic field strengths. As magnetic field strength increases, there is a substantial enhancement of both shear modulus and melting temperature. The shear modulus increases up to four times for the field strength $B^*=5000$ compared to the field-free case. The melting temperature become as high as 0.6 MeV as opposed to $\sim 0.25$ MeV (0.01 MeV$=$ $1.16 \cross 10^8$K) resulting in considerable increment. The increase in the $\mu$ and $T_m$ is a consequence of the increase in the neutron drip density with a magnetic field that allows more neutron-rich nuclei to exist at lower pressure and density.  The results suggest that magnetic field might have profound implications in the transport, cooling, and magneto-rotational evolution of a neutron star as the outer crust structure principally drives them \cite{Fatima_2020}.
\begin{figure}
    \centering
    \includegraphics[scale=0.5]{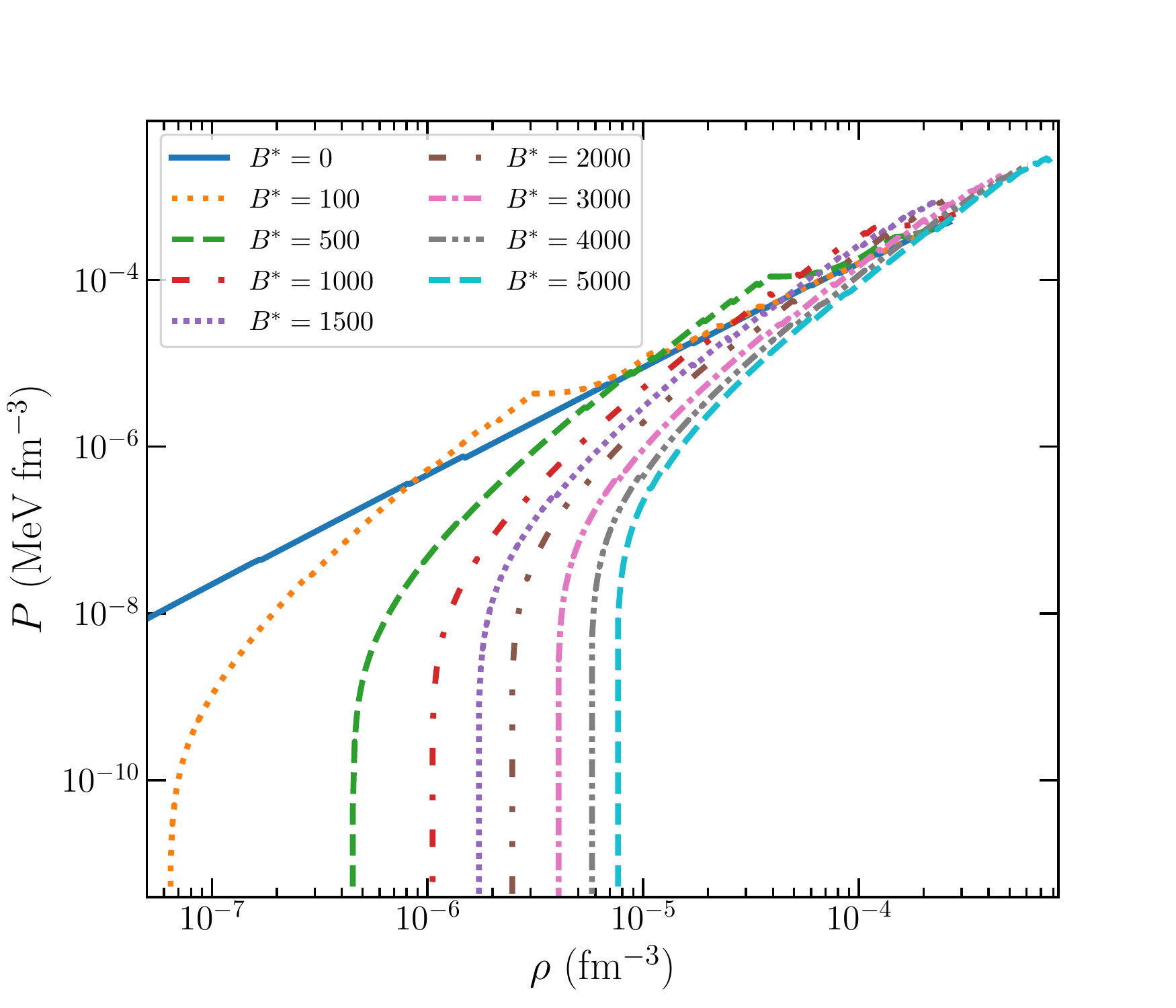}
    \caption{Equation of state of the outer crust of a neutron star at various magnetic field strengths. The theoretical mass model is taken as HFB26 with experimental evaluation of AME2020}
    \label{fig:eos_oc}
\end{figure}

In Fig. \ref{fig:eos_oc}, we show the EoS of the outer crust of a cold nonaccreting neutron star at various magnetic field strengths using the HFB26 mass model. The EoS is significantly affected in the outer crust's shallower regions, where only a few Landau levels of electrons are filled. The density of the outer crust remains unchanged for a wide range of pressure, making the matter almost incompressible in the layers adjacent to the surface of the star. The composition in this region is essentially determined by the experimental evaluation and remains model-independent. As density grows, the EoS becomes similar to the field-free case due to the rapidly filling of the Landau levels. The discontinuity in the EoS for weaker field strength signifies that the lowest Landau level is fully occupied.
\subsection{Inner crust}
\begin{figure*}
    \centering
    \includegraphics[scale=0.35]{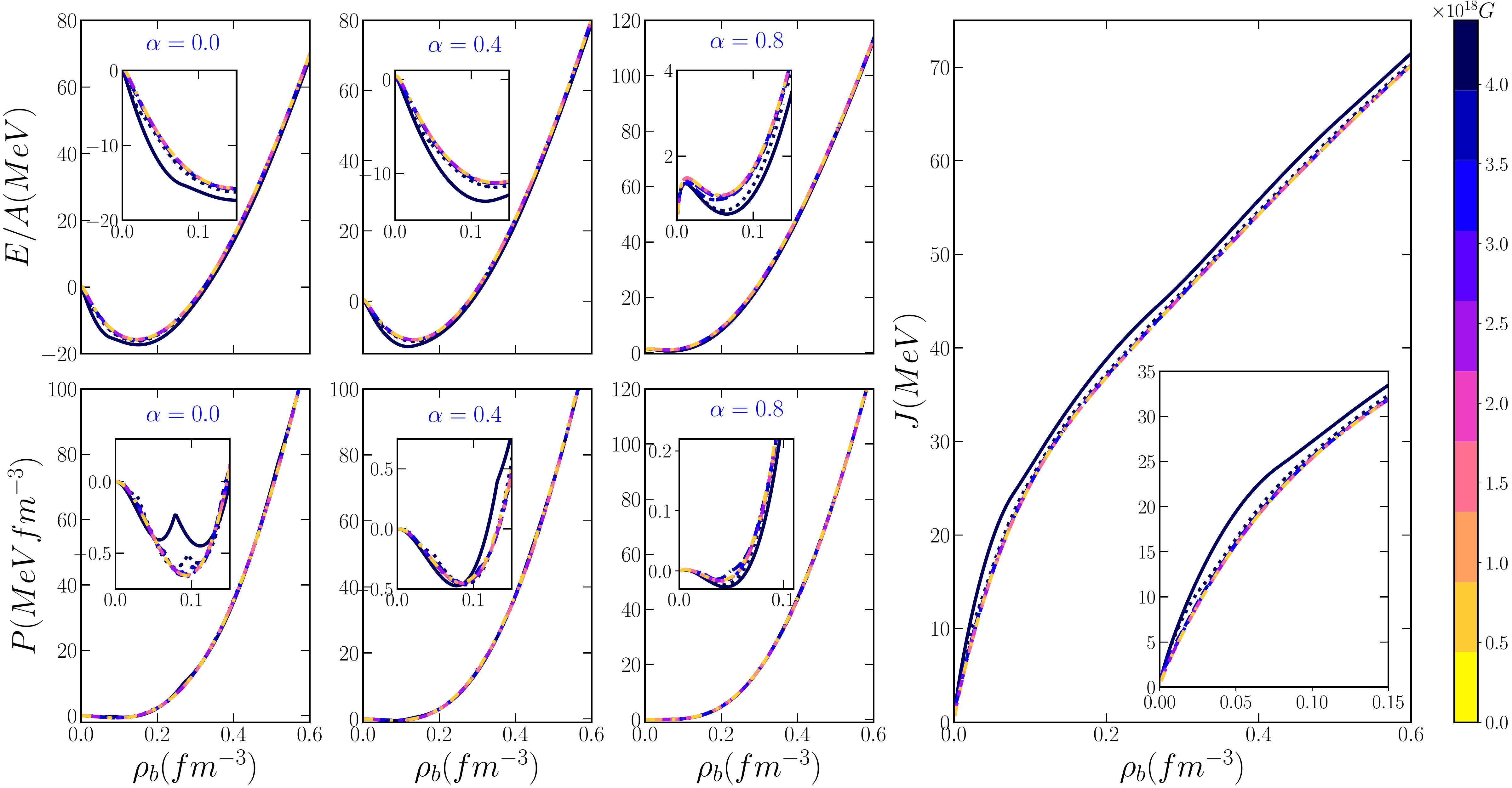}
    \caption{The variation of the binding energy ($E/A$), pressure ($P$) and symmetry energy ($J$) of homogeneous nuclear matter as a function of baryon density for various values of isospin asymmetry $\alpha=\frac{\rho_n-\rho_p}{\rho_n+\rho_p}$ at various magnetic field strengths (shown in the color bar). The parameter set taken is G3 \cite{Kumar_2017}}
    \label{fig:bulk_eos}
\end{figure*}

The calculation of the inner crust in the presence of a magnetic field is known to be tricky in the literature. Quantum oscillations possess various challenges to numerical solutions, especially in models where the energy minima becomes flat \cite{Pearson_2020, Mutafchieva_2019}. Furthermore, the small energy difference between various possible shapes (pasta structures) adds to the difficulty \cite{Parmar_2022_2}. Therefore, we extend the simplistic CLDM formalism \cite{Parmar_2022_1, Carreau_2019, Newton_2013} of the inner crust to the finite magnetic field strength. We consider the WS cell to consist of a nuclear cluster in a BCC crystal surrounded by an ultra-relativistic electron gas along with the free neutron gas.  We do not take the effect of anomalous magnetic moment (AMM) in our calculations. The AMM for electron  is insignificant in comparison to the nucleon mass \cite{Broderick_2000}, whereas for baryons, it is observed that the AMM becomes important only for $B \ge 10^{18}$G \cite{Rabhi_2015, Rabhi_2008, Lima_2013, Bao_2021}. In addition, using the one-loop fermion self-energy, Ferrer {\it et al.} \cite{Ferrer_2015} suggest that AMM of charged particles do not affect the EoS significantly. However, AMM is an important factor for higher magnetic field strengths and play a significant role in  phenomena such as axion production and neutrino-antineutrino pair emission \cite{MARUYAMA2020135413, MARUYAMA2018160}.
\begin{table}[b]
\centering
\caption{Symmetry energy ($J$) and slope parameter ($L$) coefficient for the G3 \cite{Kumar_2017}, IOPB-I \cite{Kumar_2018}, FSUGarnet \cite{Chen_2015}, IUFSU \cite{Fattoyev_2010}, IUFSU$^*$ \cite{Fattoyev_2010}  and SINPB \cite{Mondal_2016} parameter sets at saturation density ($\rho_0$) and sub-saturation density (0.05 fm$^{-3}$)}
\label{tab:rmf_sym}
\scalebox{1.0}{
\begin{tabular}{lllllll}
\hline
\hline
& FSUGarnet  & IUFSU  & IUFSU* & G3  & IOPB-I  & SINPB   \\
\hline    
$J_{\rho_0}$ (MeV) & \hfill 30.95   & \hfill 31.30  &  \hfill 29.85 & \hfill 31.84  & \hfill 33.30   &  \hfill 33.95 \\
$J_{0.05}$ (MeV) & \hfill 18.07   & \hfill 17.8 &  \hfill 15.73 & \hfill 15.66  & \hfill 15.6  &  \hfill 14.98 \\
$L_{\rho_0}$ (MeV) & \hfill 51.04   & \hfill 47.21&  \hfill 51.50 & \hfill 49.31   & \hfill 63.58  &  \hfill  71.55\\
$L_{0.05}$ (MeV) & \hfill 32.1058  & \hfill33.85  &  \hfill 32.26 & \hfill 36.781   & \hfill 37.2   &  \hfill 36.7  \\
\hline
\hline
\end{tabular}%
}
\end{table}

For the calculation of inner crust composition, We use six E-RMF parameter sets for which the value of their symmetry energy and slope parameter are given in Table  \ref{tab:rmf_sym} at saturation density and sub-saturation density (0.05 fm$^{-3}$), relevant for the inner crust. In Fig. \ref{fig:bulk_eos},  we show the variation of the binding energy $\big(\frac{E}{A}\big)$ and pressure as a function of baryon density ($\rho_b$) for various values of isospin--asymmetry at different magnetic field strength which is shown as a color bar. In addition, we also show the variation of the density-dependent symmetry energy. These calculations are performed for the parameter set G3 \cite{Kumar_2017}. The qualitative behavior of other parameters remains the same, and their behavior for the unmagnetized case has been well documented in the literature. It is clear from Fig. \ref{fig:bulk_eos} that hadronic EoS is not significantly affected for B $<10^{17}G$, which is the case of this study. The small changes appear in the subsaturation density region, which is essential for the neutron star crust. At ultra-strong magnetic field strengths, the binding energy increases, making the system more and more bound. The  variation is more pronounced for symmetric matter as more charged particles are in the system. The kinks on the pressure (especially at low density)  appear due to the successive filling of Landau levels, which disappear at high densities because of the filling of more and more  Landau levels. This behavior is analogous to free proton gas in a magnetic field \cite{Strickland_2012}.

The symmetry energy also does not change much for B $<10^{17}G$  and increases for very strong magnetic field strengths. Other nuclear matter properties, such as higher-order symmetry energy and incompressibility derivatives, also show no significant changes for B $<10^{17}G$. Therefore, the changes in the properties of the neutron star crust for the magnetic field case are predominantly due to the changes in electronic EoS, which in turn, changes the equilibrium compositions through the condition of $\beta$ equilibrium. However, we include the magnetic field on the hadronic EoS for the consistent and realistic calculations of various observables, which is absent in many crust calculations in literature  \cite{Nandi_2011}. Further, since the nuclear saturation energy does not change significantly until $B \sim 10^{18}$ G \cite{Rabhi_2015}, we use the field free value of surface energy parameters (Eqs. \ref{surf} and \ref{curv}). These fits play a crucial role in the inner crust calculations using the CLDM formalism.  For details on the surface energy fits, please see \cite{Parmar_2022_1, Parmar_2022_1, Carreau_2019}.

Out of various bulk properties of nuclear matter, symmetry energy predominantly governs the properties of asymmetric systems, which include phase transition of asymmetric nuclear matter, crustal properties of the neutron star, neutron skin thickness, etc. \cite{Vishal_2021, Dutra_2021, Parmar_2022_1, Parmar_2022_2}. In the context of neutron star crust, where the typical baryon density is in the subsaturation region,  the uncertainties in the symmetry energy account for the variation in the crustal properties. The symmetry energy is not unique to a given E-RMF parameter set and is mostly governed by the  cross-coupling of $\rho$ and $\omega$  meson. Additionally, the behavior of pure neutron matter (PNM) in the low-density regime is a crucial aspect of crustal properties, as the free neutron gas in the inner crust impacts the crust composition. The cross-coupling of the $\omega-\rho$ meson influences the PNM properties in the subsaturation density region \cite{Kumar_2017}, making it the most important factor determining the crust structure of a neutron star within the E-RMF framework. The E-RMF parameter sets used here are in harmony with the results obtained by various microscopic calculations for the PNM \cite{baldo_2008, Lovato_2022}. Furthermore, the six parameter sets differ in the way they behave in the low-density regime \cite{Parmar_2022_2} and, therefore, provide us with the flexibility to investigate the model dependency of our result and modification, if any, as compared to the zero magnetic field strength.


\begin{figure}
    \centering
    \includegraphics[scale=0.45]{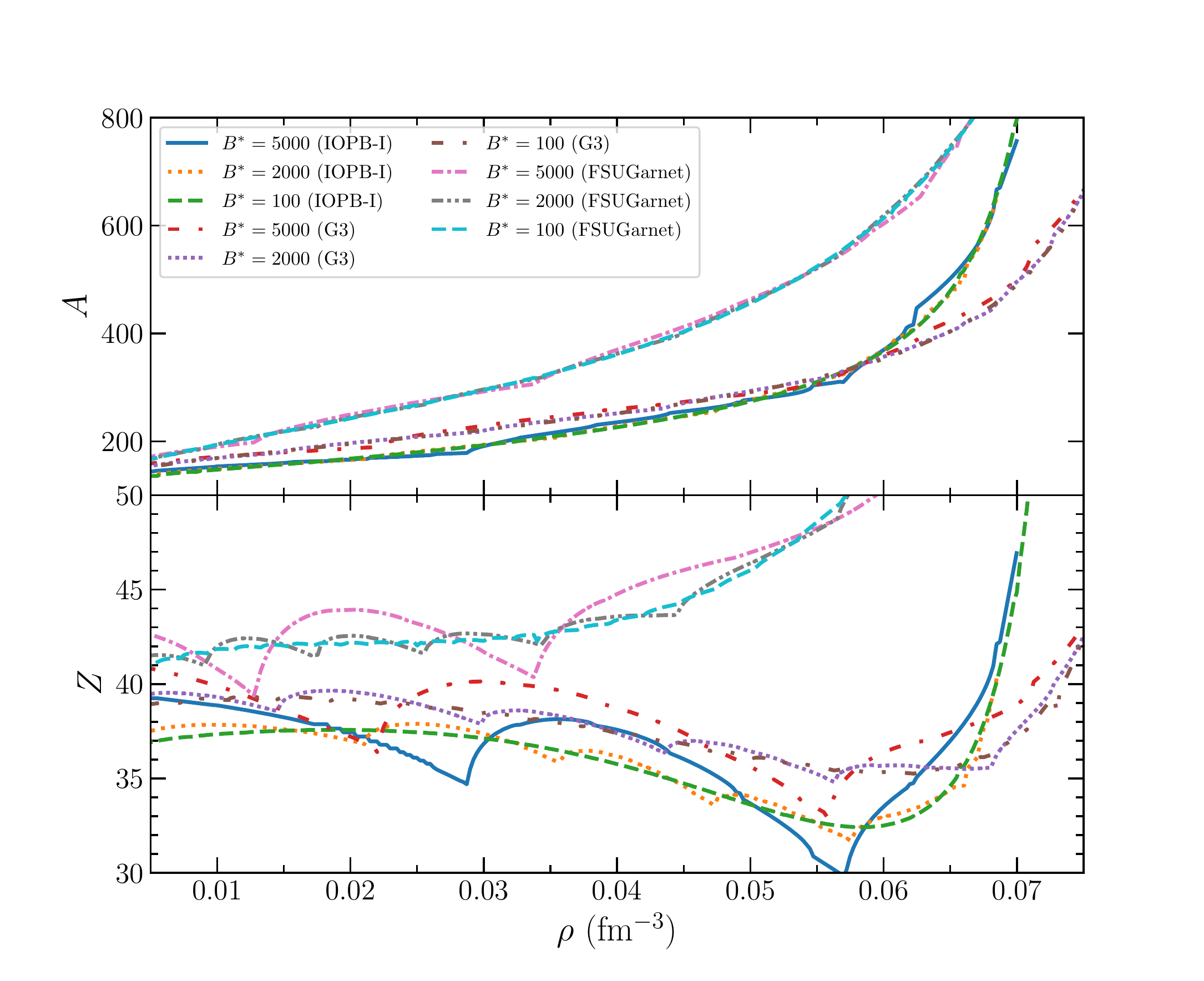}
    \caption{The distribution of the number of nucleons ($A$) in the cluster and charge number ($Z$) as a function of density in the inner crust of cold nonaccreting neutron star}
    \label{fig:inner_crust_az}
\end{figure}

In Fig. \ref{fig:inner_crust_az}, we show the distribution of the total number of nucleons in the cluster and charge number as a function of density in the inner crust of a magnetized neutron star employing the CLDM formalism. For weak magnetic field strength, the atomic number distribution is similar to the field-free case (Fig. 4 \cite{Parmar_2022_1}) as a large number of Landau levels are filled for electrons as well as protons. As the magnetic field strength increases, the atomic number distribution shows an oscillatory pattern due to the successive filling of electron Landau levels, also known as the De Haas–van Alphen effect \cite{chatterjee_2011}. For high magnetic field strength ($B^*=5000$), which becomes strongly quantizing for electrons, the oscillations typically represent the filling of $\nu=0,1,2$ Landau levels of electrons. However, the small oscillations in between the larger ones occur due to the filling of Landau levels of protons. Such behavior is similar to one obtained in Lima {\it et al.} \cite{Lima_2013} for the Thomas-Fermi calculations (see Fig. 8 and 12 of Ref. \cite{Lima_2013}).

Furthermore, the effect of symmetry energy is also evident. The oscillations become broader for the IOPB-I parameter set with the least symmetry energy at sub-saturation regions compared to the FSUGarnet with the highest symmetry energy. The difference occurs as nuclear matter EoS is affected due to the variation in electron chemical potential depending on corresponding  symmetry energy. The number of nucleons in the cluster does not change significantly for the magnetic field as high as $B^*=5000$ or $B=2.207 \cross 10^{17}$G. However, the oscillatory behavior is similar to the distribution of atomic number. 
\begin{figure}
    \centering
    \includegraphics[scale=0.5]{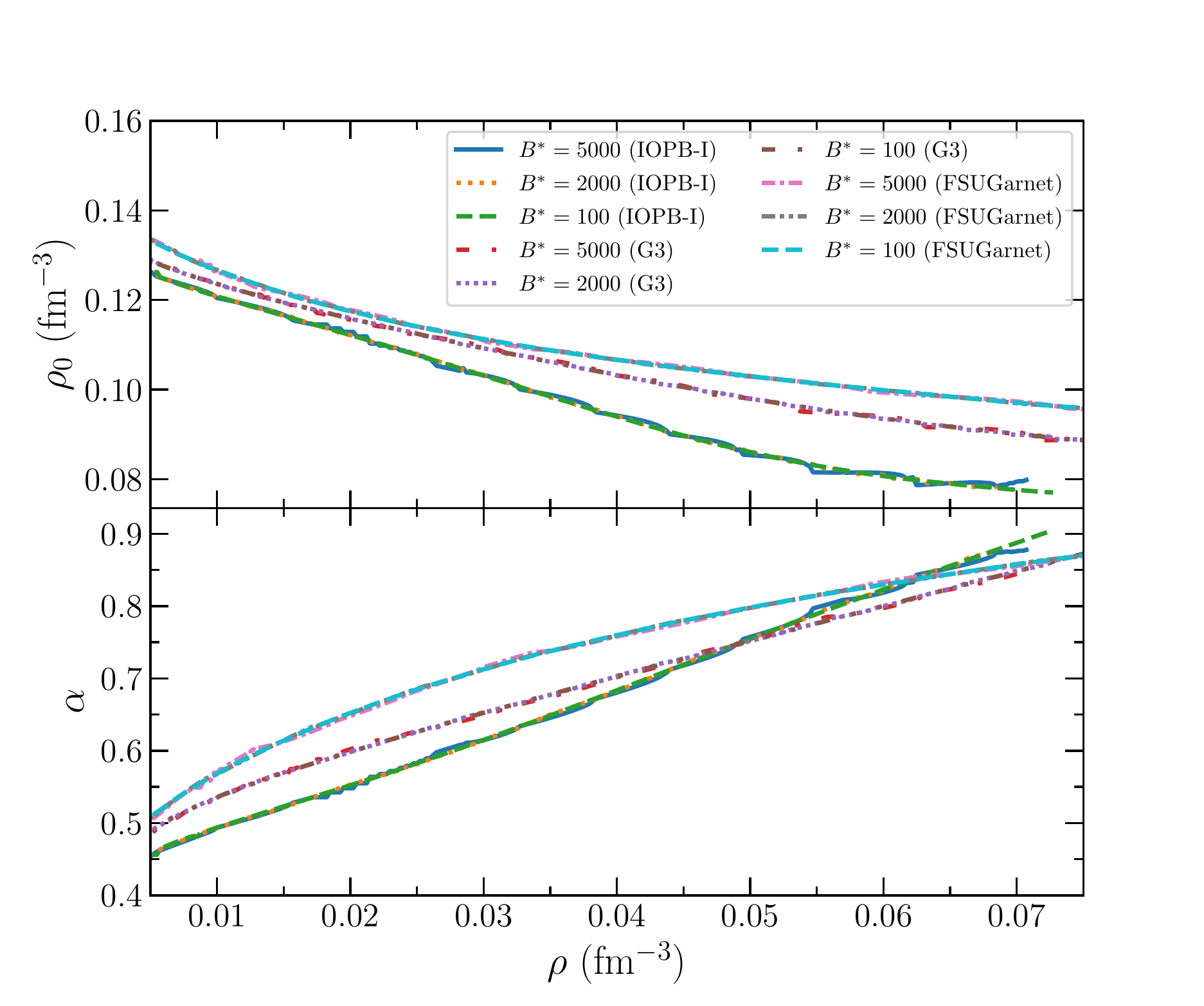}
    \caption{The average cluster density ($\rho_0$) and asymmetry in the cluster as a function of density in the inner crust for various values of magnetic field strength.}
    \label{fig:inner_crust_asy}
\end{figure}

The average cluster density and the asymmetry ($\alpha$) in the cluster are shown in Fig. \ref{fig:inner_crust_asy} for various strengths of magnetic field. The average cluster density decreases while the asymmetry increases monotonically with inner crust density for all the magnetic field strengths, making the clusters more and more dilute and neutron-rich. The magnetic field seems to have a feeble impact on the cluster density and its asymmetry, as it does not change much except for the quantum oscillations arising due to the filling of Landau levels. A closer analysis, however, reveals that in the shallower regions ($\rho_0 \le 0.01$ fm$^{-3}$) of the inner crust, which is significantly affected due to magnetic field (only $\nu=0$ level is filled), the density of the cluster is larger for the higher magnetic field strength. The cluster becomes more symmetric with an increasing magnetic field in this density range. These results agree with Ref. \cite{Mutafchieva_2019}, which uses the extended Thomas Fermi method taking the magnetic field effects only on the electrons. For $\rho>0.01$ fm$^{-3}$, the behavior of the cluster density and its asymmetry becomes comparable to the field-free case. The cluster cell size $r_c$  does not change much for the field strength $B^* \le 5000$. However, in the shallower region, it reduces as compared to the field-free case. Furthermore, the equilibrium composition of the inner crust changes significantly at very high  magnetic field strength, i.e., $B^* \ge 10000$. The presence of such a high magnetic field in the crust of magnetars has not yet been observed  \cite{Debarati_2015, Konstantinos_2016, DEXHEIMER2017487}. 

\begin{figure}
    \centering
    \includegraphics[scale=0.45]{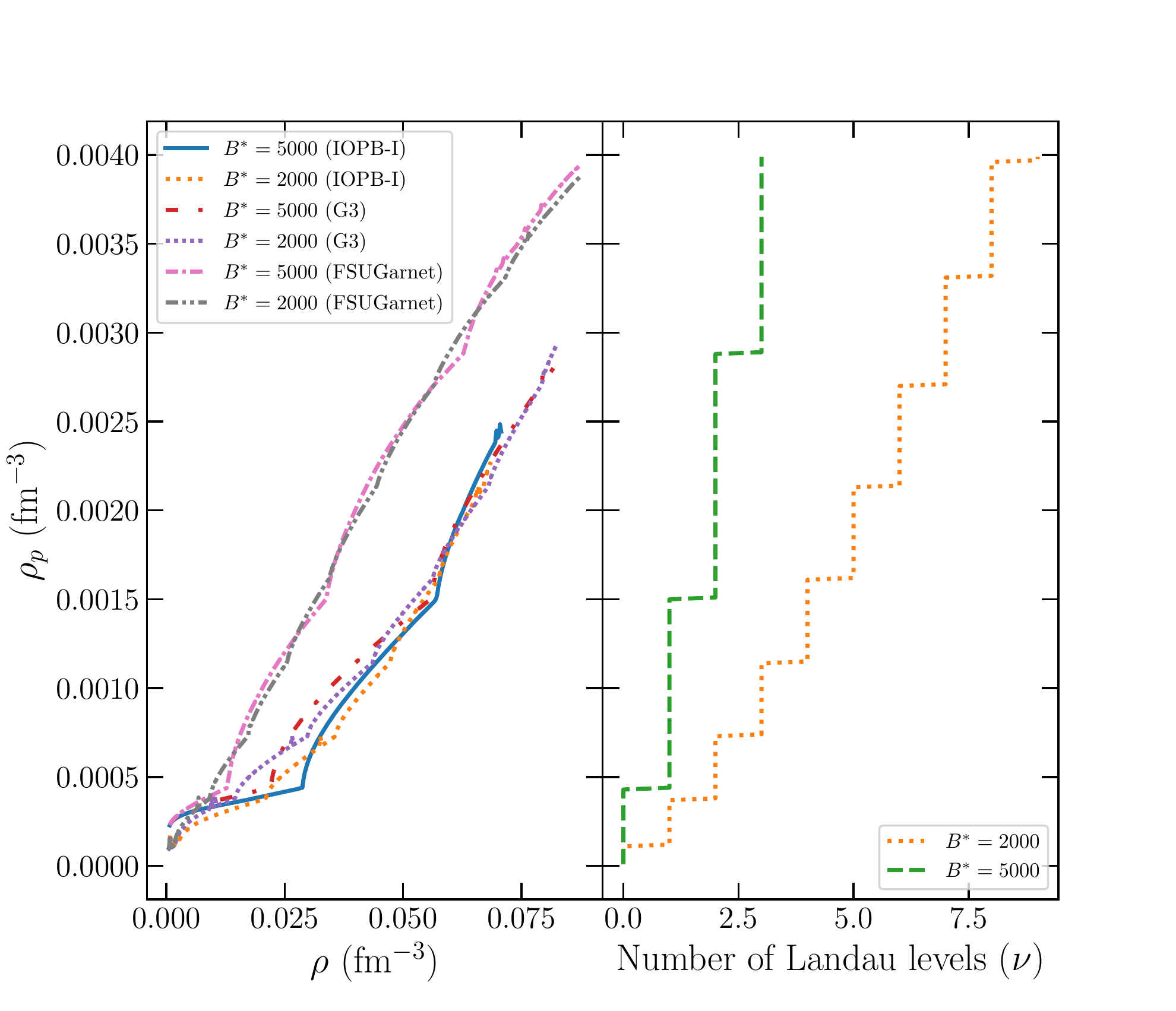}
    \caption{{\it Left:} Proton or electron density as a function of density in the inner crust. {\it Right:} The number of filled Landau levels of electrons at a given proton or electron density.}
    \label{fig:inner_crust_pd}
\end{figure}
From the above analysis, it is clear that the magnetic field causes quantum oscillations in the inner crust composition, where electrons play  more crucial role as compared to  the baryons. For baryons, the critical magnetic field \Big($B^P_c=\big(\frac{M_p}{M_e}\big)^2B^e_c=1.487 \cross 10^{20}$ G\Big) is substantially greater than that of electrons. Since the magnetic field considered in this work and those observed in the neutron star's crust is much lower than the $B^P_c$, it does not significantly affect baryon properties. We show the proton density, which is equal to electron density in the inner crust of a neutron star in Fig. \ref{fig:inner_crust_pd} along with the number of Landau levels filled by electrons. The proton density increases monotonically with the inner crust density. The equilibrium proton density at a given  density in the inner crust depends on the parameter set used. The FSUGarnet with the largest symmetry energy in the subsaturation region estimates a higher proton density than the G3 and IOPB-I sets, with lower symmetry energy at a particular magnetic field. However, the fluctuations in the proton density are guided by the filling of electron Landau levels. The discontinuity in the proton density arises where the filling of the subsequent Landau level takes place. This discontinuity occurs for the same proton density for all the parameter sets but at different inner crust densities.  We further observe in our calculations that the density of neutron gas is not significantly affected by the presence of the magnetic field. 
\begin{figure}
    \centering
    \includegraphics[scale=0.5]{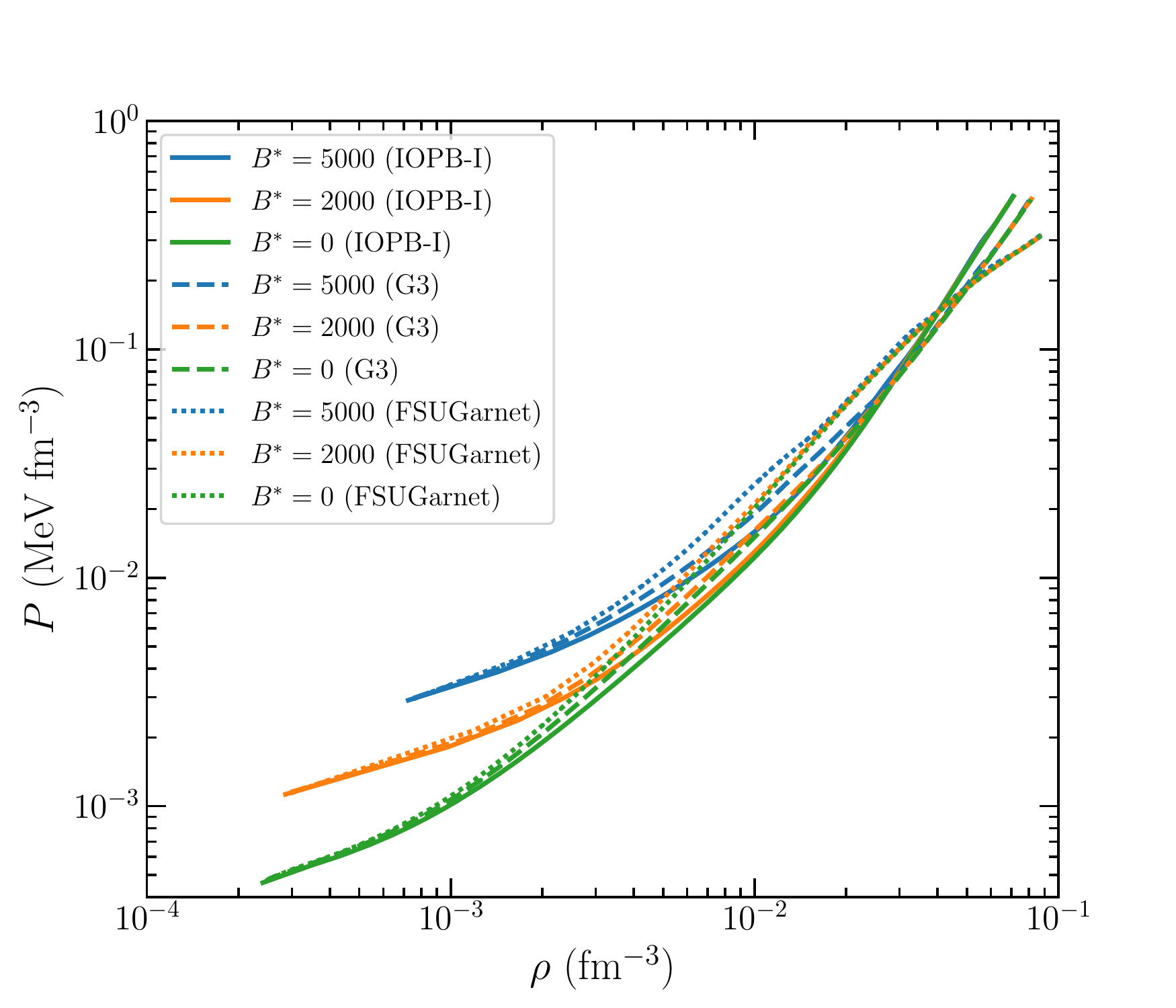}
    \caption{Equation of state (EoS) of the inner crust of a magnetized neutron star using  IOPB-I, G3, and FSUGarnet E-RMF parameter sets.}
    \label{fig:inner_crust_eos}
\end{figure}

The EoS of the inner crust at various magnetic fields for the E-RMF parameter sets FSUGarnet, G3, and IOPB-I are shown in Fig. \ref{fig:inner_crust_eos}. We also show the EoS for the field-free case for comparison. It is seen that the magnetic field effects become prominent in the lower-density regions of the inner crust. As density approaches the crust-core transition density, magnetic field effects vanish as several Landau levels of electron and protons are filled, and EoSs imitates the field-free case. The inner crust experiences higher pressure at lower density for the larger magnetic field strength. The EoS preserves its dependence on the symmetry energy as for a given magnetic\label{ic} field strength, FSUGarnet shows the stiffest EoS followed by the G3 and IOPB-I parameter sets, a trend observed for the field free case \cite{Parmar_2022_1}. We also compare the results of EoS with switching on/off the magnetic field effects on the baryons and observe that the electrons play a far  critical role in the inner crust calculations than the protons. Furthermore, the above calculations are also performed for the IUFSU, IUFSU$^*$, and SINPB parameter sets, and it is seen that the qualitative results remain the same. 
\begin{table}
\centering
\caption{The density (fm$^{-3}$) of the onset of pasta structures in the inner crust of magnetized neutron star using the  G3 \cite{Kumar_2017}, IOPB-I \cite{Kumar_2018}, FSUGarnet \cite{Chen_2015}, IUFSU \cite{Fattoyev_2010}, IUFSU$^*$ \cite{Fattoyev_2010}  and SINPB \cite{Mondal_2016} parameter sets. Hom. represents the crust-core transition density.}
\label{tab:pasta_density}
\scalebox{1.0}{
\begin{tabular}{lllllll}
\hline
\hline
                           &       & \multicolumn{5}{c}{Density (fm$^{-3}$)}    \\
                           \cline{3-7}
                           & B*    & Rod    & Slab   & Tube   & Bubble & Hom.    \\
\hline
\multirow{5}{*}{FSUGarnet} & 10000 & 0.0443 & 0.0583 & 0.0812 & 0.0857 & 0.0913 \\
                           & 5000  & 0.0453 & 0.0590 & 0.0811 & 0.0864 & 0.0925 \\
                           & 3000  & 0.0456 & 0.0591 & 0.0811 & 0.0864 & 0.0926 \\
                           & 1000  & 0.0456 & 0.0591 & 0.0810 & 0.0865 & 0.0926 \\
                           & 0     & 0.0456 & 0.0590 & 0.0810 & 0.0865 & 0.0918 \\
&&&&&&\\
\multirow{5}{*}{IUFSU}     & 10000 & 0.0483 & 0.0634 & 0.0850 & 0.0896 & 0.0945 \\
                           & 5000  & 0.0497 & 0.0644 & 0.0846 & 0.0902 & 0.0960 \\
                           & 3000  & 0.0498 & 0.0643 & 0.0847 & 0.0900 & 0.0966 \\
                           & 1000  & 0.0500 & 0.0642 & 0.0847 & 0.0901 & 0.0965 \\
                           & 0     & 0.0499 & 0.0641 & 0.0847 & 0.0901 & 0.0965 \\
&&&&&&\\                           
\multirow{5}{*}{IUFSU*}    & 10000 & 0.0529 & 0.0632 & 0.0779 & 0.0824 & 0.0855 \\
                           & 5000  & 0.0526 & 0.0647 & 0.0807 & 0.0837 & 0.0855 \\
                           & 3000  & 0.0522 & 0.0652 & 0.0804 & 0.0836 & 0.0856 \\
                           & 1000  & 0.0526 & 0.0650 & 0.0807 & 0.0836 & 0.0858 \\
                           & 0     & 0.0525 & 0.0652 & 0.0807 & 0.0836 & 0.0858 \\
&&&&&&\\
\multirow{5}{*}{G3}        & 10000 & 0.0564 & 0.0631 & 0.0796 & 0.0834 & 0.0885 \\
                           & 5000  & 0.0559 & 0.0648 & 0.0833 & 0.0858 & 0.0896 \\
                           & 3000  & 0.0546 & 0.0657 & 0.0825 & 0.0854 & 0.0894 \\
                           & 1000  & 0.0549 & 0.0655 & 0.0830 & 0.0853 & 0.0889 \\
                           & 0     & 0.0551 & 0.0655 & 0.0830 & 0.0853 & 0.0889 \\
&&&&&&\\
\multirow{5}{*}{IOPB-I}    & 10000 & 0.0548 & 0.0615 &        &        & 0.0712 \\
                           & 5000  & 0.0545 & 0.0617 &        &        & 0.0743 \\
                           & 3000  & 0.0542 & 0.0620 &        &        & 0.0737 \\
                           & 1000  & 0.0541 & 0.0619 &        &        & 0.0736 \\
                           & 0     & 0.0542 & 0.0618 &        &        & 0.0735 \\
&&&&&&\\
\multirow{5}{*}{SINPB}     & 10000 & 0.0499 & 0.0573 &        &        & 0.0600 \\
                           & 5000  & 0.0479 & 0.0542 &        &        & 0.0623 \\
                           & 3000  & 0.0488 & 0.0545 &        &        & 0.0620 \\
                           & 1000  & 0.0490 & 0.0542 &        &        & 0.0613 \\
                           & 0     & 0.0490 & 0.0543 &        &        & 0.0611\\
\hline
\hline

\end{tabular}%
}
\end{table}

It has been previously shown and verified in our calculations that with an increase in magnetic field strength, the system's free energy decreases due to the Landau quantization \cite{Broderick_2000, Lima_2013}. Therefore, it may  impact the appearance of non-spherical shapes in the higher-density regime of the inner crust. To investigate this, we show the onset density of different pasta structures in the inner crust of a magnetized neutron star in Table \ref{tab:pasta_density} for various magnetic field strengths. The density at which different pasta structures appear does not change much for  $B^*=1000$. For $B^*>1000$, the onset densities changes, however, not significantly except for the crust-core transition density. These changes or small fluctuations in the onset density for different magnetic field strengths are not monotonic and essentially depend on the magnetic field strength and EoS. These results align with the self-consistent Thomas-Fermi approximation of Bao {\it et al.} \cite{Bao_2021}. The substantial changes can only be seen for magnetic field strength as high as $B^*=10000 = 4.414\cross10^{17}$ G. For $B^*=10000$, and the crust-core transition density decreases compared to the field-free case as the free energy decreases faster at higher density for higher magnetic fields. We do not see any change in the number of pasta structures for a given E-RMF parameter set as compared to the field-free case, even for the magnetic field strength of $B^*=10000$.

The feeble changes in the pasta onset density and considerable change in the pressure because of the magnetic field effect in the inner crust prompted us to investigate the modifications which might occur in the mass and the thickness of pasta layers. The mass and thickness of the pasta are sensitively affected by the pressure and chemical potential as given by \cite{Newton_2021},
\begin{equation}
\label{eq:rr}
    \frac{\Delta R_p}{\Delta R_c} \approx \frac{\mu_c-\mu_p}{\mu_c-\mu_0},
\end{equation}
\begin{equation}
\label{eq:pp}
    \frac{\Delta M_p}{\Delta M_c} \approx 1-\frac{P_p}{P_c}.
\end{equation}
 Here, $\mu_c$, $\mu_p$, and $\mu_0$ are the baryon chemical potential at crust-core (CC) transition, the location at which the pasta structure starts, and at the surface of the star. $P_p$ and $P_c$ are the pressure at the bottom of the pasta layer and at the crust-core transition density.  
\begin{figure*}
  \centering
\subfloat[]{%
  \label{fig:pasta_mass}
  \includegraphics[scale=0.45]{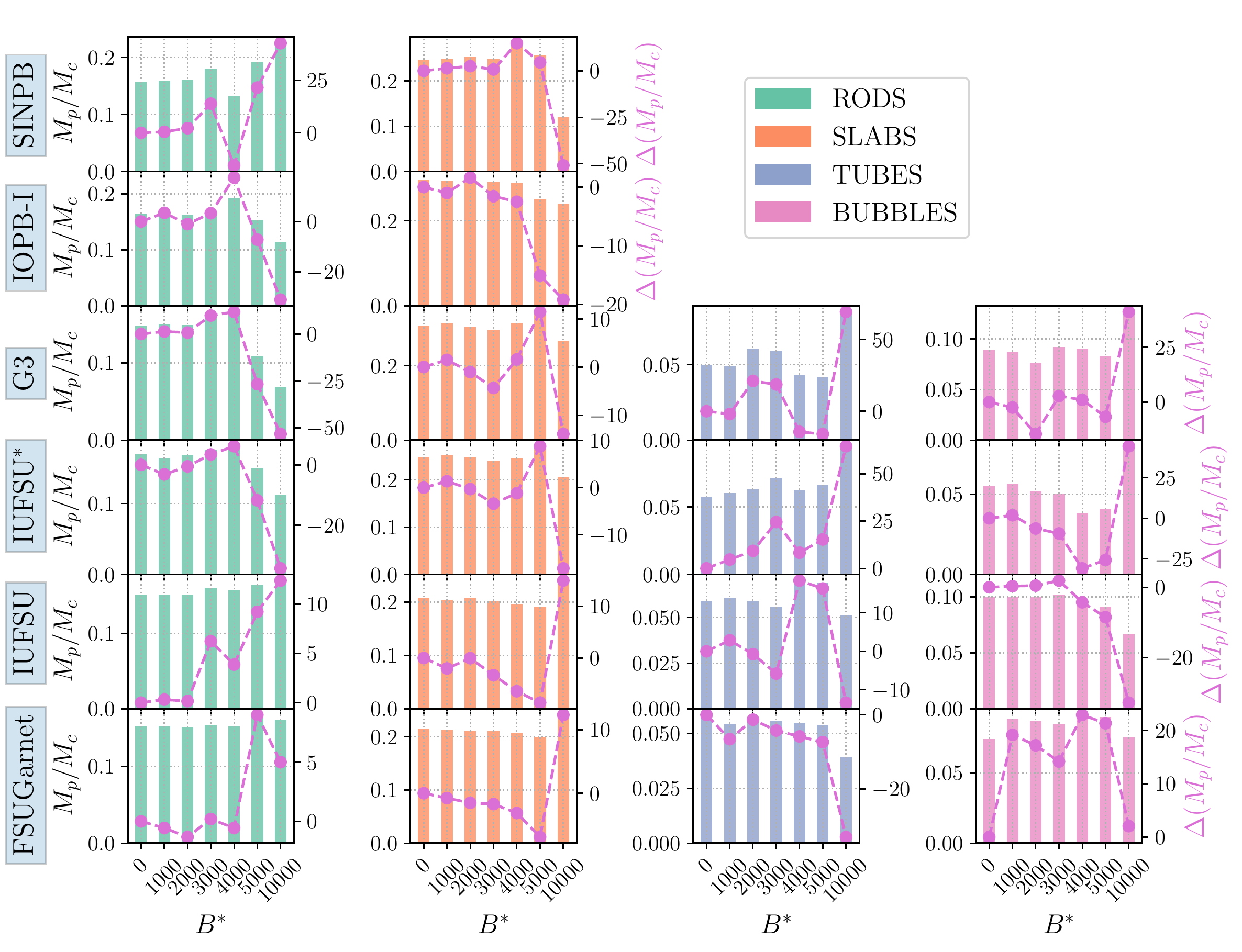}%
}\hfill
\subfloat[]{%
  \label{fig:pasta_thick}
  \includegraphics[scale=0.45]{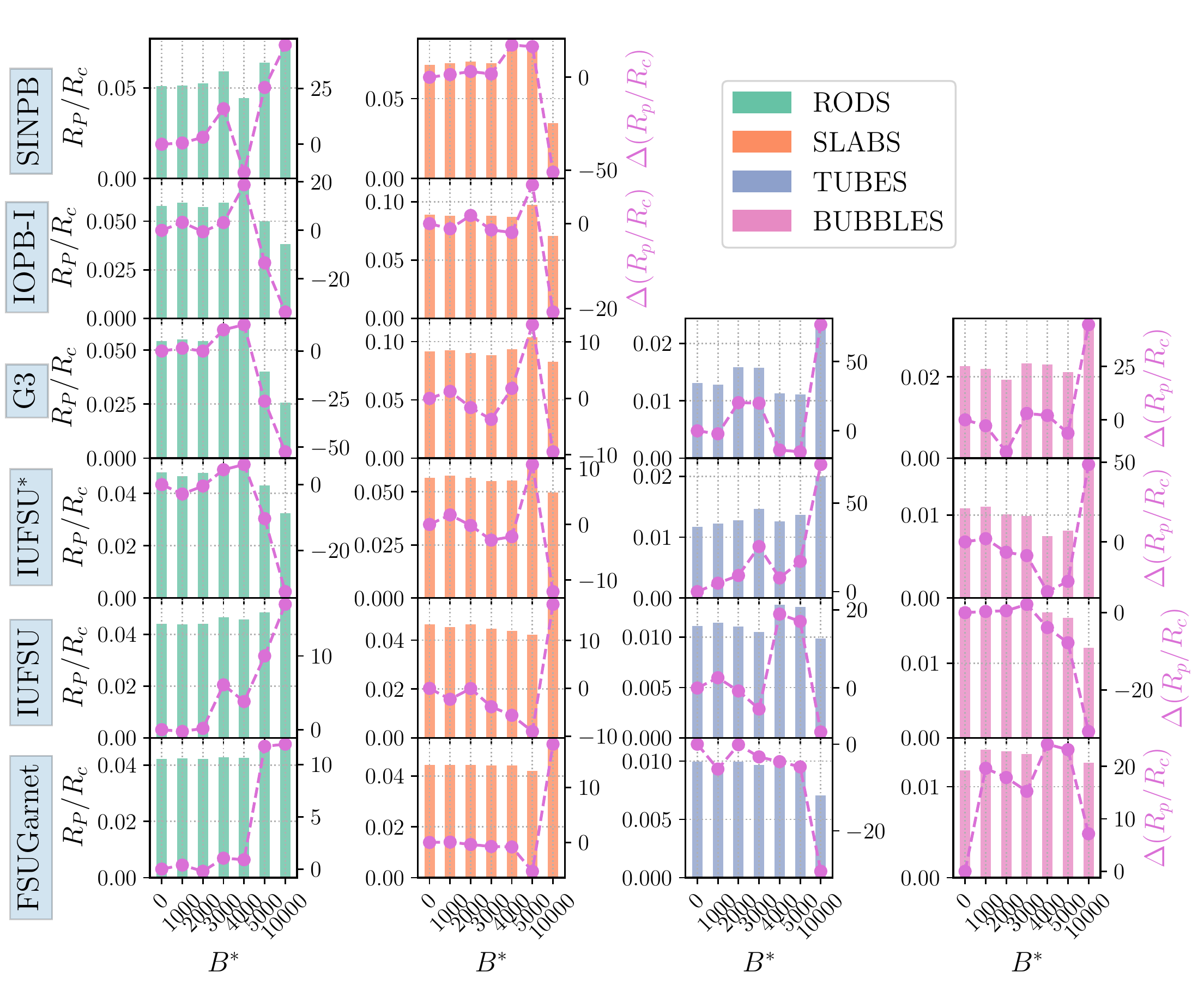}%
} 
\caption{(a) The relative mass and (b) thickness of the different layers of pasta structures at various magnetic field strengths for FSUGarnet, IUFSU, IUFSU$^*$, G3, IOPB-I, and SINPB E-RMF parameter sets. The left scale shows the absolute values, while the secondary right scale represents the percentage change with respect to the field-free case. }
\end{figure*}

\begin{figure*}
    \centering
    \includegraphics[scale=0.6]{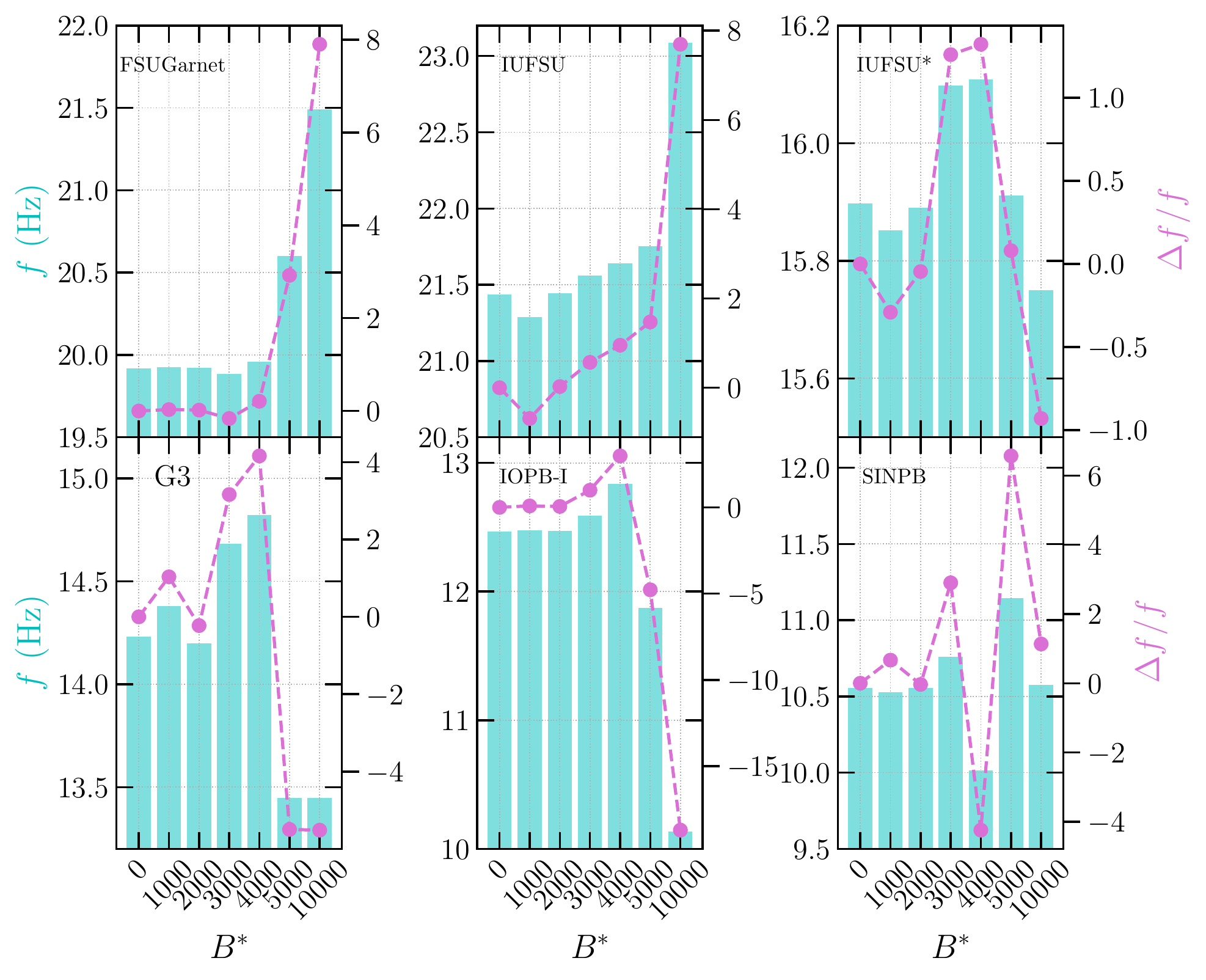}
    \caption{Frequency of fundamental torsional oscillation mode ($l=2$) in the crust of a magnetized neutron star for maximum mass at various magnetic field strengths.}
    \label{fig:freq}
\end{figure*}

We show the relative mass and thickness of different pasta structures, namely; rods, slabs, tubes, and bubbles, in Fig. \ref{fig:pasta_mass} and \ref{fig:pasta_thick} for the  FSUGarnet, IUFSU, IUFSU$^*$, G3, IOPB-I, and SINPB parameter sets. We take the highest field strength $B^*=10000$ to investigate the thickness and mass for the upper limit. The E-RMF sets are plotted with increasing values of symmetry energy in the sub-saturation density regions, with FSUGarnet having the highest symmetry energy and SINB the least. The relative mass and thickness of the different pasta layers are sensitive to a) the strength of the magnetic field and b) the EoS used. The changes with respect to the field-free case are similar to the mass and thickness of a particular pasta layer at a particular strength of the magnetic field. It is apparent that the relative mass and thickness of  pasta layers  fluctuate with the magnetic field  and do not behave in a particular fashion. One can see that deviations as high as ($25-30$) \% with respect to the field free case can be seen for magnetic field strength $B^*=10000$ while it remains $\sim (10-15)\ \%$ for $B^*=5000$. There is an interesting behavior of EoS on the relative thickness and mass of pasta layers. The gross trend of fluctuations for a particular pasta layer changes when we compare different E-RMF calculations. The trend which is shown by E-RMF parameter sets having larger symmetry energy in the subsaturation density regions, namely: FSUGarnet and IUFSU, get reversed for higher magnetic field strength in comparison to IUFSU$^*$, G3, IOPB-I, and SINPB having comparatively lower symmetry energy. The large fluctuations have a profound implication for a neutron star  cooling period, where the magnetic field's decay occurs. The inner crust structure, therefore, can change considerably with the time period and magnetic field strength.

One of the significant characteristics of the inner crust is its shear modulus and shear speed which plays a crucial role in the crustal physics of neutron stars. We calculate these quantities using the E-RMF forces considered in the above analysis using the same method as in Ref. \cite{Parmar_2022_1} and observe that they also experience typical fluctuation due to the Landau quantization. Considering the pasta phases to have no shear modulus as prescribed in \cite{Parmar_2022_2}, we calculate the frequency of the fundamental torsional oscillation mode using the approximate solution of crustal shear perturbation equation as \cite{Gearheart_2011, Samuelsson_2006, Piro_2005}
\begin{equation}
\label{eq:freq}
    \omega_0^2 \approx \frac{e^{2\nu} V_s^2 (l-1)(l+2)}{2RR_c},
\end{equation}
where $e^{2\nu}= 1-2M/R$, $R$ is the radius of the star, $R_c$ is the radius of the crust and  $l$  is the angular `quantum' number. In Fig. \ref{fig:freq}, we show the frequency variation of fundamental torsional oscillation mode ($l = 2$) in the crust of a magnetized neutron star using various E-RMF forces. We consider no entrainment effects in these calculations; therefore, these estimates can be considered lower bound on the fundamental torsional frequency \cite{Gearheart_2011}. Further, the shear modulus is calculated at the boundary of non-spherical structures in the inner crust. Out of all the forces, only FSUGarnet and IUFSU satisfy the possible candidate for the fundamental mode of QPOs: 18 Hz and 26 Hz \cite{Gearheart_2011}. The fundamental frequency also oscillates like other crustal properties of neutron stars with changing magnetic fields. For E-RMF force having a higher value of symmetry energy in the subsaturation region, the frequency tends to increase with the magnetic field, where it increases as high as 8\% as compared to the field-free case. Parameter sets IUFSU$^*$, G3 and IOPB-I also estimate an increase in frequency; however, a sharp dip for field strength higher than $B^*>4000$ can be seen for these sets. The frequency varies as low as 15 \% for the IOPB-I set. For SINPB having the least symmetry energy, more prominent fluctuations are observed. Therefore, the torsional oscillation mode frequency is sensitive to both the magnetic field strength and EoS (symmetry energy of the EoS in the subsaturtion region). These changes are significant in context to QPO, which are major asteroseismological sources to constrain neutron star crust properties such as pasta structures.

It may be noted that we do not consider possible deformation of the Wigner–Seitz(WS) cell due to the presence of magnetic field in this work. In literature, such attempts have are primarily based on Thomas-Fermi (TF) approximation with the modification that deformation of the cylindrical nature is introduced in the electronic distribution with spherically symmetric or deformed WS cell \cite{nag2010crustal, ghosh2011theoretical}. In such studies, a predetermined nature of deformation of the WS cell and the electronic gas distribution (cylindrical, prolate, etc.) is necessary, which might not be true always. The highly sophisticated molecular dynamics simulations of the present day can be used for such analysis. Moreover, solving Poisson's equations in the cylindrical coordinate need a lot of approximation to handle the boundary value problems, such as using only the lowest Landau level, etc., and require a lot of computational power \cite{Avancini_2008}.

\subsection{Unified EoS and Neutron star observables}
\label{unified}
\begin{figure}
    \centering
    \includegraphics[scale=0.5]{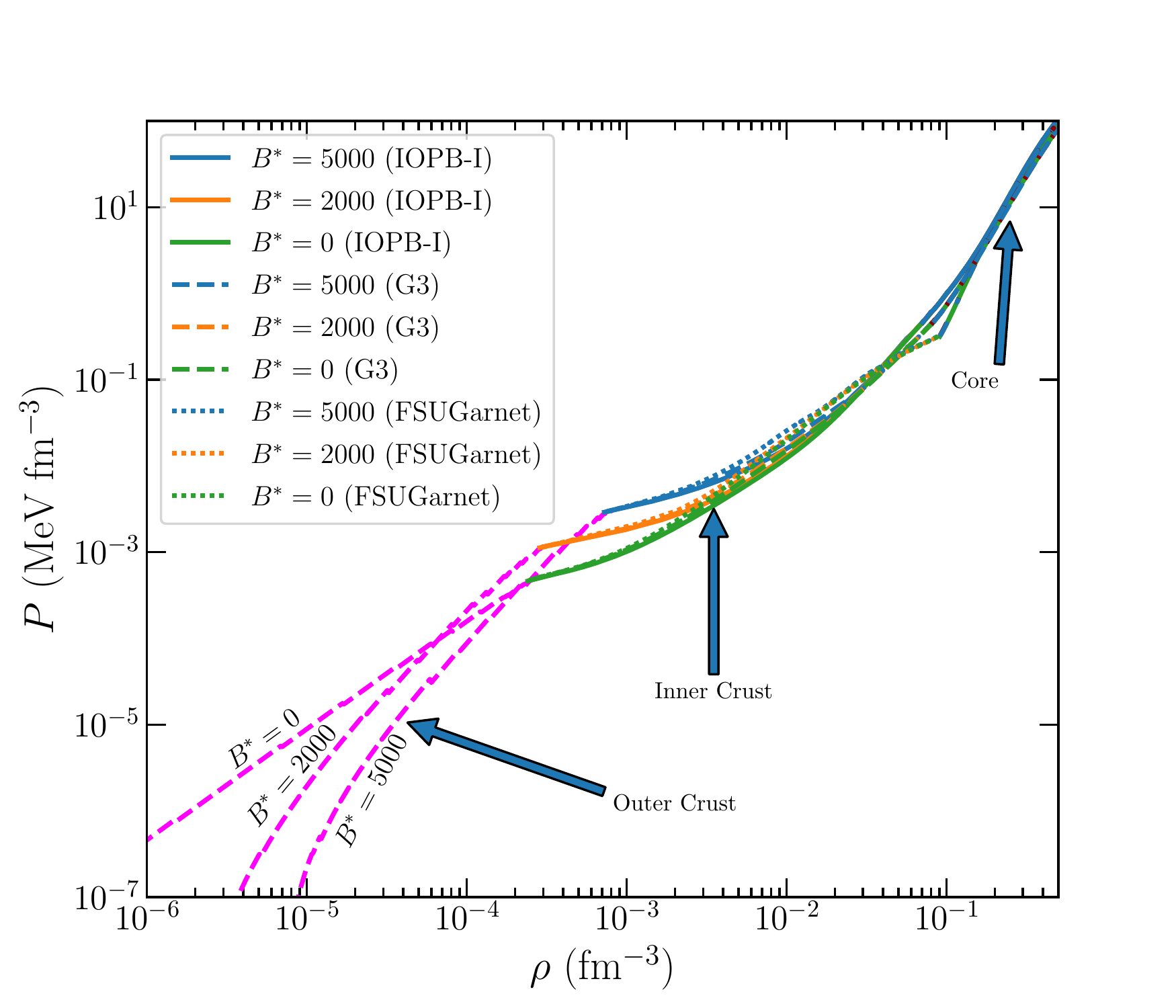}
    \caption{Unified Equation of state (EoS) for a magnetized neutron star at various magnetic field strengths.}
    \label{fig:unifid_eos}
\end{figure}
For completeness, we model a magnetized neutron star in a unified way from the surface to the core under the condition of $\beta$ equilibrium and charge neutrality. The core consists of neutrons, protons, electrons, and muons. The outer and inner crust EoS have been defined in the earlier section, whereas the EoS of the core is estimated using the same E-RMF set for which the inner crust is calculated. This procedure ensures the consistency between the crust and the core and helps to estimate the crustal parameters better. In general, the  density \cite{Debades_1997}, or chemical potential \cite{DEXHEIMER2017487} dependent magnetic field variation, are used for the neutron star calculation in literature. However, we approximate the same magnetic field strength throughout the neutron star to calculate the properties such as outer and inner crust thickness, mass, etc. This approximation is taken for the simplicity of the calculations. It can be considered equivalent to the density/chemical potential dependent magnetic field profiles where the central field is quite low \cite{Patra_2020, Rather_2021}. In such circumstances, the EoS does not deviate much from the field-free case, and the magnetic field almost becomes constant for most parts of NS \cite{Patra_2020, Rather_2021}. 

We show the unified EoS using the FSUGarnet, G3, and IOPB-I  parameters sets for a magnetized neutron star in Fig. \ref{fig:unifid_eos}. The magnetic field effect become profound in the outer crust and shallow regions of the inner crust, whereas it vanishes in the core. The reason can be attributed to more Landau level filling at a large density, which makes the EoS similar to the field-free case.  A significant variation in the outer and inner crust EoS with changing magnetic fields means that the crust structure will be affected significantly, consequently impacting various microscopic properties of a neutron star such as elastic properties, cooling, bursts, etc. These conclusions remain valid even for more realistic magnetic field profiles. 
\begin{table*}
\centering
\caption{The neutron star properties for canonical mass ($M=1.4M_\odot$), corresponding radius ($R_{\rm 1.4}$), radius of the outer crust ($R_{\rm oc}$), radius of the total crust ($R_{\rm c}$), mass of the outer crust ($M_{\rm oc}$), mass of the total crust ($M_{\rm crust}$),
normalized maximum MI ($I_{\rm 1.4}$), fractional moment of inertia for outer crust ( $I_{ \rm oc}/I_{ \rm 1.4}$ ) ,fractional moment of inertia for total crust ( $I_{\rm c}/I_{\rm 1.4}$ ),   second Love number ($k_{2,1.4}$) and dimensionless tidal deformability ($\Lambda_{\rm max}$)  for various EoSs at various magnetic field strength.}
\label{tab:NS_properties}
\renewcommand{\tabcolsep}{0.2cm}
\renewcommand{\arraystretch}{1}
\begin{tabular}{llllllllllll}
\hline
\hline
\begin{tabular}[c]{@{}l@{}}Parameter\\ sets\end{tabular}         & B* & \begin{tabular}[c]{@{}l@{}}$R_{1.4}$\\ (km)\end{tabular}  & \begin{tabular}[c]{@{}l@{}}$R_{oc}$\\ (km)\end{tabular} & \begin{tabular}[c]{@{}l@{}}$R_{c}$\\ (km)\end{tabular} & \begin{tabular}[c]{@{}l@{}}$M_{\rm oc}$\\ ($M_\odot$)\end{tabular} & \begin{tabular}[c]{@{}l@{}}$M_{\rm c}$\\ ($M_\odot$)\end{tabular} & $I_{1.4}$  & $I_{oc}/I_{1.4}$ & $I_c/I_{1.4}$ & $k_{2,1.4}$  & $\lambda_{1.4}$ \\
\hline

\multirow{3}{*}{FSUGarnet} & 0  & 13.1935 & 0.5291   & 1.35600   & 0.00005   & 0.02473   & 0.34297 & 0.00007      & 0.03630      & 0.08891 & 629.131  \\
                        & 2000 & 12.9951 & 0.5269 & 1.20570 & 0.00011 & 0.02391 & 0.35245 & 0.00017 & 0.03456 & 0.09480 & 619.817 \\
                        & 5000 & 12.9881 & 0.5274 & 1.19900 & 0.00028 & 0.02424 & 0.35292 & 0.00044 & 0.03506 & 0.09514 & 620.265 \\
                        &      &          &         &         &         &         &         &         &         &         &           \\
\multirow{3}{*}{IUFSU}  & 0    & 12.6075 & 0.4728 & 1.17940 & 0.00004 & 0.02392 & 0.35138 & 0.00005 & 0.03507 & 0.08704 & 488.953 \\
                        & 2000 & 12.6153 & 0.4903 & 1.16400 & 0.00010 & 0.02563 & 0.35271 & 0.00015 & 0.03744 & 0.08844 & 498.451 \\
                        & 5000 & 12.6092 & 0.4906 & 1.15680 & 0.00025 & 0.02574 & 0.35314 & 0.00039 & 0.03764 & 0.08873 & 498.730 \\
                        &      &          &         &         &         &         &         &         &         &         &           \\
\multirow{3}{*}{IUFSU$^*$} & 0    & 12.9221 & 0.5025 & 1.23540 & 0.00004 & 0.02563 & 0.34704 & 0.00006 & 0.03780 & 0.08843 & 563.083 \\
                        & 2000 & 12.8100 & 0.5089 & 1.12460 & 0.00010 & 0.02605 & 0.35308 & 0.00016 & 0.03813 & 0.09230 & 561.581 \\
                        & 5000 & 12.8039 & 0.5098 & 1.12070 & 0.00027 & 0.02658 & 0.35350 & 0.00042 & 0.03891 & 0.09259 & 562.141 \\
                        &      &          &         &         &         &         &         &         &         &         &           \\
\multirow{3}{*}{G3}     & 0    & 12.6278 & 0.4753 & 1.26130 & 0.00004 & 0.03376 & 0.34453 & 0.00005 & 0.05035 & 0.08128 & 460.236 \\
                        & 2000 & 12.4931 & 0.4785 & 1.15580 & 0.00009 & 0.03257 & 0.35175 & 0.00014 & 0.04810 & 0.08546 & 458.863 \\
                        & 5000 & 12.4864 & 0.4800 & 1.15200 & 0.00024 & 0.03313 & 0.35220 & 0.00038 & 0.04895 & 0.08576 & 459.176 \\
                        &      &          &         &         &         &         &         &         &         &         &           \\
\multirow{3}{*}{IOPB-I} & 0    & 13.3316 & 0.5420 & 1.35420 & 0.00005 & 0.04004 & 0.34420 & 0.00007 & 0.05862 & 0.09244 & 686.640 \\
                        & 2000 & 13.3117 & 0.5580 & 1.33620 & 0.00012 & 0.03954 & 0.34491 & 0.00019 & 0.05791 & 0.09281 & 684.730 \\
                        & 5000 & 13.2987 & 0.5583 & 1.32950 & 0.00031 & 0.04055 & 0.34567 & 0.00050 & 0.05938 & 0.09335 & 685.145 \\
                        &      &          &         &         &         &         &         &         &         &         &           \\
\multirow{3}{*}{SINPB}  & 0    & 13.1576 & 0.5250 & 1.18490 & 0.00005 & 0.03053 & 0.34245 & 0.00006 & 0.04608 & 0.08814 & 612.943 \\
                        & 2000 & 13.2057 & 0.5481 & 1.23350 & 0.00012 & 0.02914 & 0.33945 & 0.00019 & 0.04401 & 0.08605 & 609.610 \\
                        & 5000 & 13.1297 & 0.5423 & 1.16630 & 0.00030 & 0.03024 & 0.34348 & 0.00048 & 0.04565 & 0.08865 & 610.223 \\
\hline
\hline
\end{tabular}%
\end{table*}

Finally, we calculate the global properties, such as mass, radius, etc., of a magnetized neutron star at a particular magnetic field using the TOV Eqs. \ref{eq:pr} and \ref{eq:mr} considering the star to be spherically symmetric. This assumption remains valid for the strength of the magnetic field considered in this work \cite{Patra_2020}. We observe that the maximum mass and radius of the star do not change significantly up to $B^*=5000$ as the core EoS remains almost unaffected. The E-RMF parameter sets considered in this work reproduce the maximum mass and radius in agreement with the ($M=2.01^{+0.04}_{-0.04} M_\odot$) and ($M=2.14^{+0.10}_{-0.09}+M_\odot$) limit \cite{Antoniadis_2013, Cromartie_2019}.  To infer the effect of magnetic field on the crust mass and thickness, we show neutron star properties for canonical mass ($M=1.4M_\odot$), corresponding radius ($R_{\rm 1.4}$), the radius of the outer crust ($R_{\rm oc}$), the radius of the total crust ($R_{\rm c}$), the mass of the outer crust ($M_{\rm oc}$), the mass of the total crust ($M_{\rm crust}$), normalized maximum MI ($I_{\rm 1.4}$), the fractional moment of inertia for the outer crust ( $I_{ \rm oc}/I_{ \rm 1.4}$ ), the fractional moment of inertia for the total crust ($I_{\rm c}/I_{\rm 1.4}$), second Love number ($k2_{1.4}$) and dimensionless tidal deformability ($\Lambda_{\rm 1.4}$) for various EoSs at various magnetic field strengths in Table \ref{tab:NS_properties}. Although the mass of the total crust remains almost similar with changing magnetic field strength, the outer crust mass increases $\sim 5-6$ times as compared to the field-free case. However, the thickness of the outer crust does not change much with magnetic field strength. The fractional crustal moment of inertia of the total crust also remains unaffected. In contrast, the outer crust moment of inertia increases $\sim 7-8$ times compared to the field-free case. The insensitiveness of the total crust mass, thickness, and crustal moment of inertia can be understood from Table \ref{tab:pasta_density} and Fig. \ref{fig:inner_crust_eos}, which shows a minute change in crust-core transition density and pressure (and consequently chemical potential ) which decides the mass and thickness of the crust. 

\section{CONCLUSION}
\label{conclusion}
In summary, we investigate the impact of the magnetic field of the order of $\sim 10^{17}$G on the neutron star crust and associated phenomena. We minimize the Gibbs free energy for the outer crust using the nuclear mass from the atomic mass evaluations AME2020 clubbed with the theoretical mass models of HFB, producing the up-to-date sequence of nuclei in the magnetized outer crust of a neutron star.   The outer crust structure predominantly depends on the structural effects in atomic nuclei and becomes increasingly symmetric as we increase the magnetic field strength. In the presence of magnetic field , more neutron-rich (less bound nuclei) can sustain at lower pressure in the outer crust due to the magnetic field. The surface density, neutron drip density, and neutron drip pressure increase substantially with an increase in the strength of the magnetic field. 

To calculate the equilibrium composition of the inner crust in the presence of the magnetic field, we use the CLDM formalism for the first time using the E-RMF framework for nuclear interaction. The equilibrium composition of the inner crust shows typical quantum fluctuations, which become prominent at a high magnetic field. In the shallow regions of the inner crust, the nuclear cluster becomes more dense and symmetric with an increase in magnetic field strength. The EoS of the electron plays the predominant role in infusing the magnetic field effects on inner crust EoS. In contrast, the symmetry energy of the nuclear matter EoS dictates the in-situ characteristics. The inner crust calculations using the CLDM formalism agree with the available self-consistent Thomas-Fermi calculations. 

We further investigate the possible change in the pasta structures in the presence of the magnetic field. The onset density of the various pasta structures fluctuates with the magnetic field strength. It results in substantial modifications in their mass and thickness, which ranges as high as 50\%. The frequency of the fundamental torsional oscillation mode is also investigated, and an increase/decrease of $\sim$ 5\% is seen depending upon the EoS and the magnetic field strength.

Finally, we construct the unified EoS of a magnetized neutron star by estimating the core EoS using the same E-RMF parameter as for the inner crust, ensuring consistency between both layers. The magnetic field impacts the outer crust and the shallower regions of the inner crust as the EoS of the electron gas becomes strongly quantizing in this region. As density increases, a large number of Landau level fillings lead to the core EoS imitating the field-free EoS. We then calculate the global properties for a canonical mass star, as the maximum mass does not change for the considered magnetic field. The outer crust mass and its fractional crustal moment of inertia increase substantially with an increasing magnetic field.  However, the total crust mass and thickness remain immune to the changing magnetic field strength. 

Although the present work considers magnetized neutron star crust to be composed of cold-catalyzed matter, it is unlikely for the crust, especially the outer crust, to be in complete thermodynamic equilibrium during its formation. Therefore, the structure of the neutron star crust in the presence of a magnetic field and some finite temperature becomes essential. This improvement deserves more investigation and will be carried out in future work to understand the neutron star structure holistically. 
\section{Acknowledgement}
We would like to thank the anonymous reviewers for the careful reading of our manuscript and their
insightful suggestions.
\bibliography{mag_field}

\end{document}